 \documentclass[final,3p,times, square, comma, sort&compress] {elsarticle}
 
\usepackage{amssymb}

\usepackage{subfig}

\usepackage{placeins}

\usepackage{afterpage}

\begin{document}

\begin{frontmatter}



 \title{ Test of scintillating bars coupled to Silicon Photomultipliers for a charged particle tracking device}

\author[rvt]{S. Cecchini}

\author[rvt]{I. D'Antone}

\author[rvt]{L. Degli Esposti}

\author[rvt]{I. Lax}

\author[rvt]{G. Mandrioli}

\author[rvt]{N. Mauri}

\author[rvt,rv]{L. Pasqualini\corref{cor1}}
\cortext[cor1]{Corresponding Author: Laura Pasqualini; Email: laura.pasqualini@bo.infn.it; Phone: +393355701960}


\author[rvt]{L. Patrizii}

\author[rvt]{M. Pozzato}

\author[rvt]{G. Sirri}

\author[rvt]{M. Tenti}

\address[rvt]{Istituto Nazionale di Fisica Nucleare, I.N.F.N., Sezione di Bologna,
Viale Berti Pichat 6/2 40127 Bologna, Italy}
\address[rv]{Dipartimento di Fisica e Astronomia, Universit\`a di Bologna,
Viale Berti Pichat 6/2 40127 Bologna, Italy}

\address{}

\begin{abstract}
The results obtained in laboratory tests, using scintillator bars read by silicon photomultipliers are reported. The present approach is the first step for designing a precision tracking system to be placed inside a free magnetized volume for the charge identification of low energy crossing particles. The devised system is demonstrated able to provide a spatial resolution better than 2 mm.

\end{abstract}


\end{frontmatter}

\section{Introduction}
\label{intro}
The main aim of WA104-NESSiE R\&D project was the development of innovative experimental solutions for the search for sterile neutrinos with a new CERN-SPS neutrino beam, as proposed in \cite{Bernardini, ICARUS}.

Among the planned activities was the construction of a light spectrometer seated in a 20-30 m$^{3}$ magnetized air volume, the Air Core Magnet (ACM). The whole design should be optimised for the determination of the momentum and charge of muons in the 0.5 - 5 GeV/c range (the mis-identification is required to be less than 3\% at 0.5 GeV/c). Monte Carlo (MC) simulations show that a tracking device of low-density material along the beam direction and a spatial resolution of  $\sim$ 1.5 mm is required inside the magnetized air volume.

In this paper we report the results obtained with a small array of triangular scintillator bars coupled to silicon photomultiplier (SiPM) with wavelength shifter (WLS) fibers.  Solid polystyrene scintillator bars are commonly employed in current and planned experiments for particle physics at accelerators  \cite{Berra,Noah,Aliaga,montanari,T2K,MINOS}.
In our test we have used the extruded scintillator bars produced  in triangular shape by FNAL \cite{FNAL}. This bar profile is here demonstrated able to provide the necessary spatial resolution in reconstructing the position of the crossing particle by profiting of the charge-sharing between adjacent bars readout in analog mode. 
SiPMs are excellent candidates in replacing standard photomultipliers in many experimental conditions. Features of SiPM like single photon detection, reduced size, low power consumption, insensitiveness to magnetic field \cite{Dolgoshein} make them a natural choice in designing a large tracking device to be placed inside a magnetized volume. 
Tests have been performed with laser beam pulses and radioactive source in order to characterize the scintillator bar response and SiPM behaviour.

An experimental set-up for detecting cosmic rays (CR) has been used in order to finalize the design and the arrangement of a multi-plane prototype for further test with accelerator beam in order to verify the performances of the detector final design. 

\section{Silicon PhotoMultipliers characterisation}\label{sec:SiPM}

Many authors have reported extensively about characterization and performances of SiPMs proposed for or adopted in high energy particle physics \cite{Berra,Noah,Aliaga,montanari} and space-born experiments \cite{sirad}. Here we briefly present the observed behaviour of the SiPM used in our tests regarding the main sources of noise and the effect of temperature on its response and linearity.

Several models and packaging have been considered. Finally the MicroSL-10035 X13 SMD SiPM provided by the SenSL \cite{sipm} manufacturer was chosen. It has an active area of about 1x1 mm$^{2}$ with 504 microcells for an overall fill factor of 64\%. The breakdown voltage, as reported in the datasheet, is (27.5 $\pm$ 0.5) V and the overvoltage V$_{bias}$ is allowed to range between 1 and 5 V.

\subsection{Dark current}

\begin{figure}[t!]
 \centering
  \includegraphics[width=0.52\textwidth]{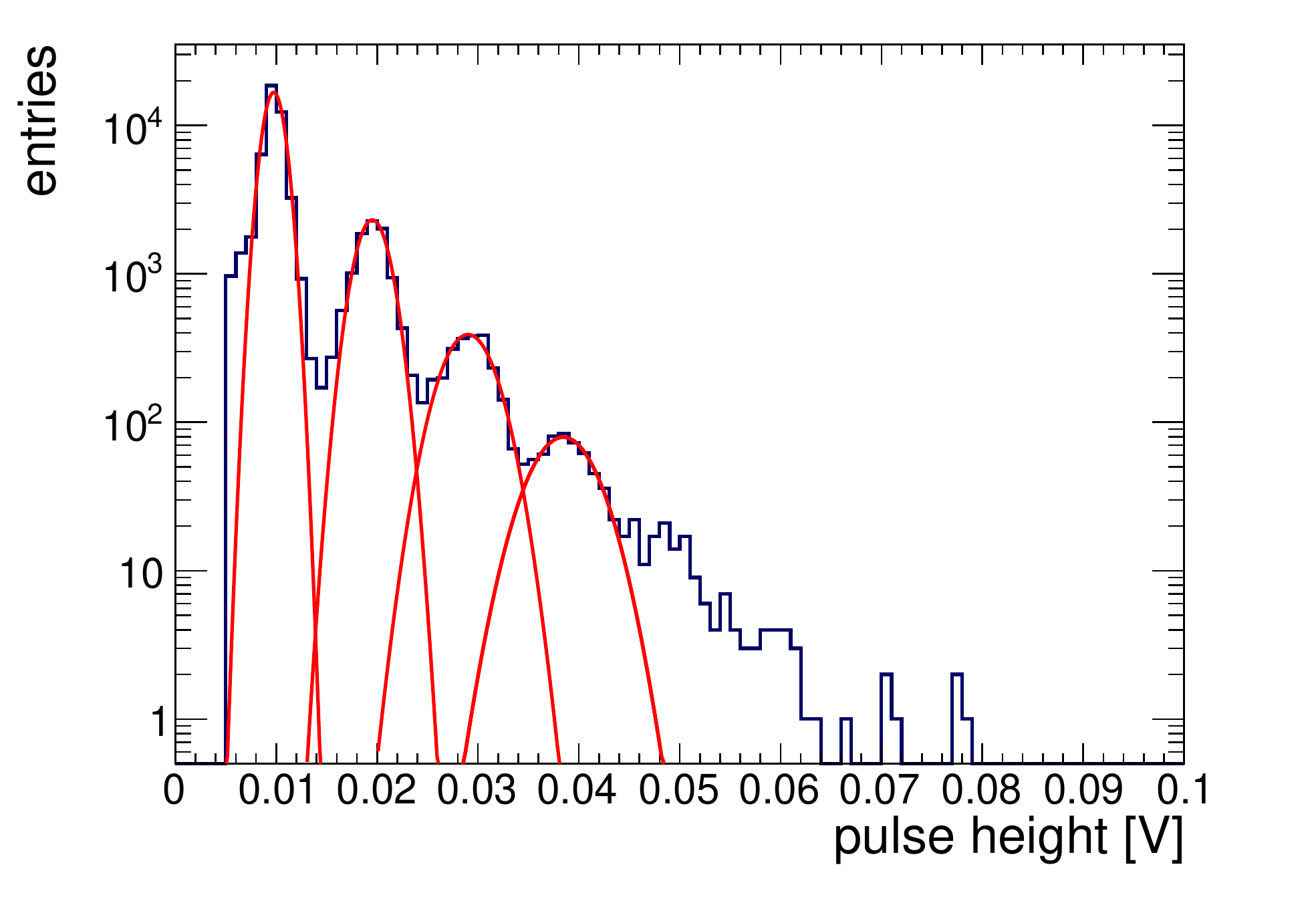}
 \caption{\footnotesize Pulse height distribution of the dark current for the MicroSL-10035 SiPM. The total collection time is 25 ms at $V_{bias}$ = 1.5 V and T = 26 $^{\circ}$C.  The dark current rate including other kind of noise (i.e. cross talk and after pulses) is estimated 900 kHz, with a cross talk probability of about 15\%. The red curves, representing Gaussian distributions, are drawn to guide the eye in identifying the peaks. The 5th and 6th peaks are also barely visible. }
\label{fig:dark}
\end{figure}

The main source of noise which limits the SiPM's single photon resolution is the ``dark current'' rate. It is originated by charge carriers thermally created in the sensitive volume and present in the conduction band and therefore it depends on the temperature. The dark current pulse height distribution, measured at working voltage V$_{wk}$ equal to 29 V and temperature about 26 $^{\circ}$C is shown in Fig. \ref{fig:dark}. 

\begin{figure}[t!]
 \centering
  \includegraphics[width=0.52\textwidth]{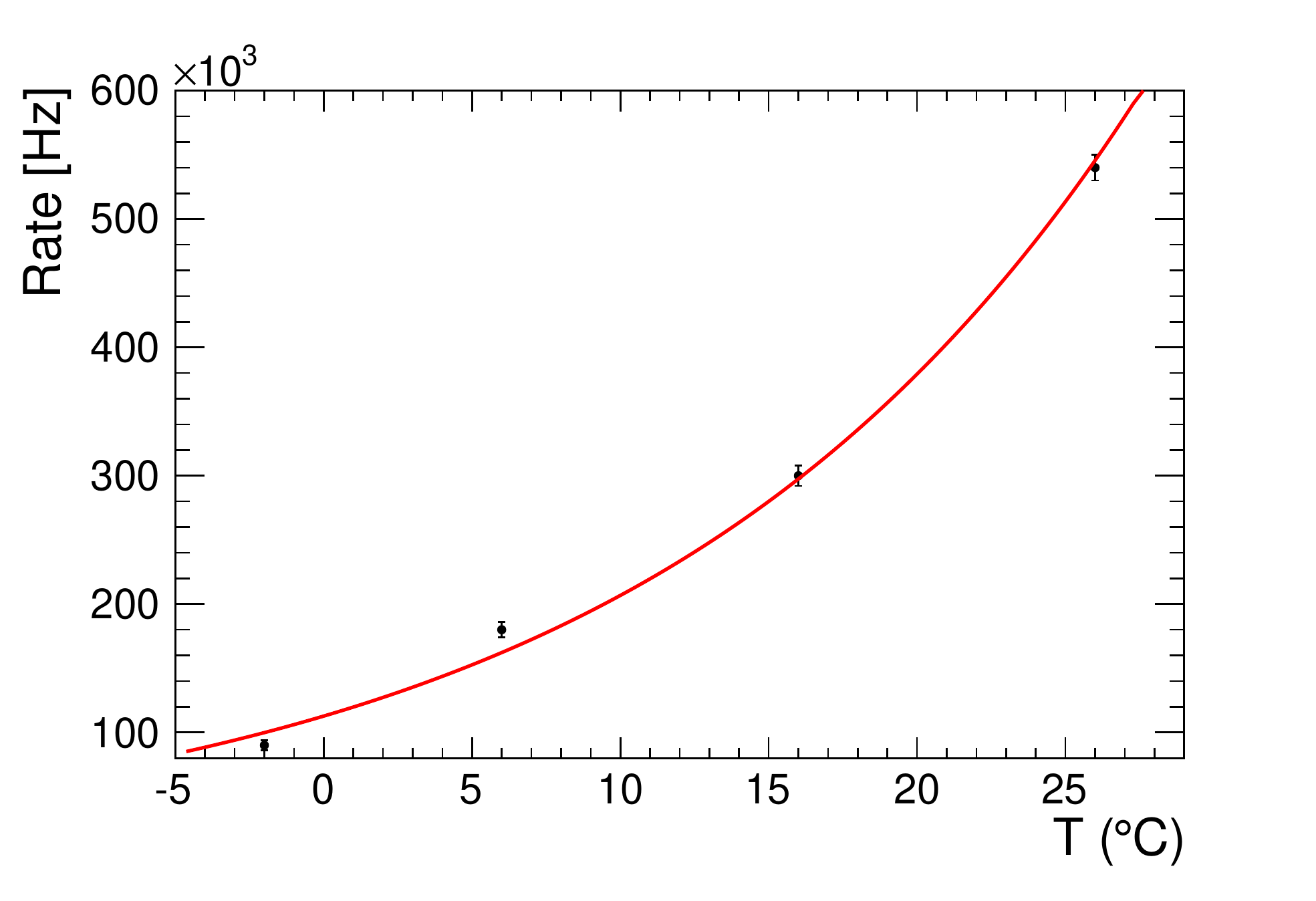}
 \caption{\footnotesize  Single pixel rate dependence on environmental temperature of the selected SiPM for fixed V$_{bias}$ = 1.5 V. The red line is drawn to guide the eye along the measured values.}
\label{fig:rate_T}
\end{figure}

The dependence of the dark current single pixel rate as a function of the temperature has been investigated using Peltier cells in order to change and keep the temperature controlled. The results are shown in Fig. \ref{fig:rate_T}. In the typical range of laboratory environment (18 $^{\circ}$C - 28 $^{\circ}$C) it is possible to estimate a variation of the dark current rate of $\sim$ 5 \%/$^{\circ}$C.

\begin{figure}[t!]
 \centering
  \includegraphics[width=0.40\textwidth]{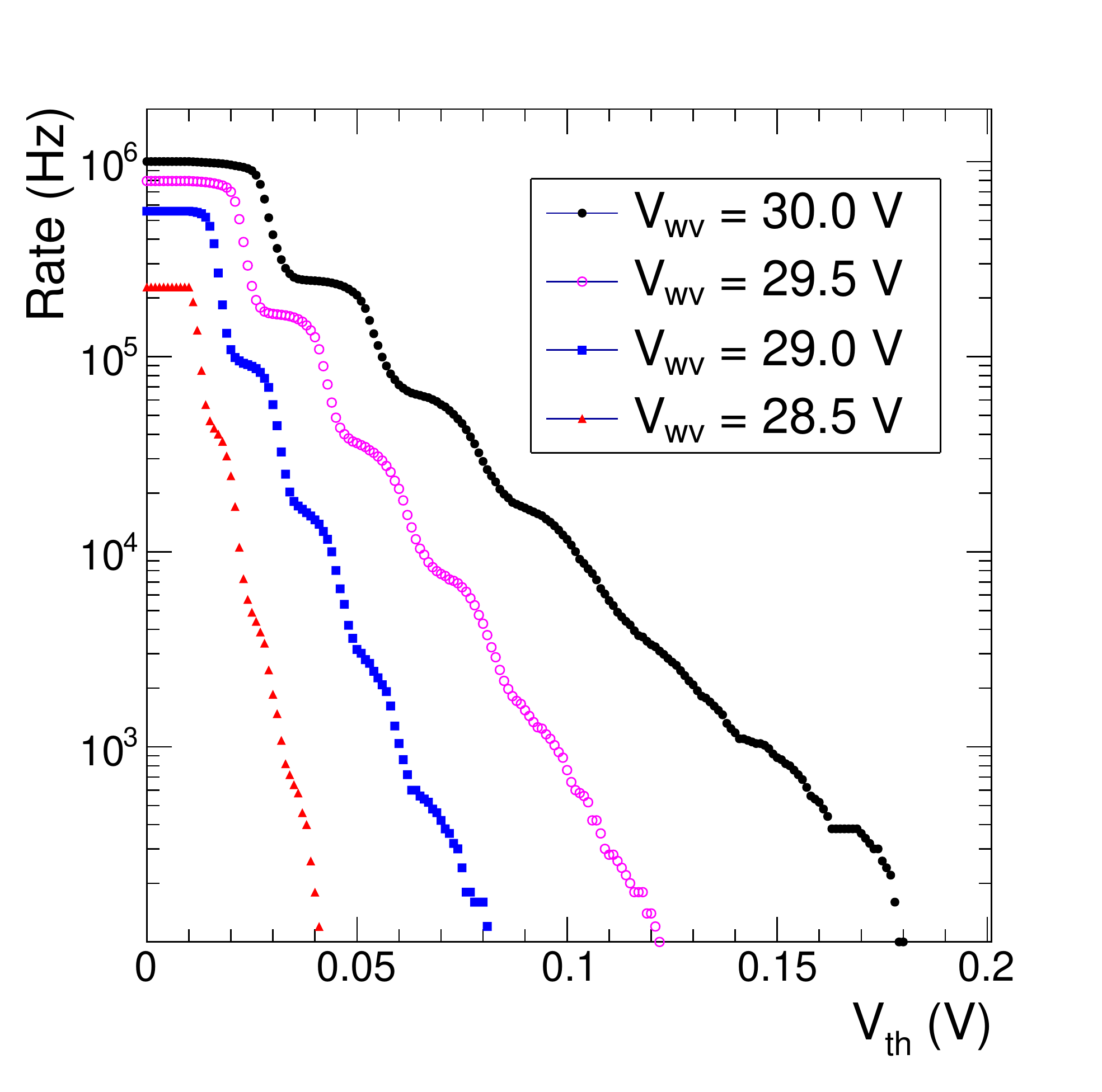}
 \caption{\footnotesize Dark current rate of the selected SiPM for different working voltages versus the threshold voltage V$_{th}$. Measurements have been performed at T = 26 $^{\circ}$C.}
\label{fig:rate_Vth}
\end{figure}

Dark current rate depends also on the V$_{wk}$ as shown in Fig. \ref{fig:rate_Vth}. In order to have low rates of dark current the value of V$_{bias}$ has been fixed at 1.5 V giving a working voltage V$_{wk}$ of 29 V. It is evident that, if necessary, it can be convenient to use a bias voltage regulator which automatically compensates for temperature variations.

Not always the pixels of the SiPM work independently from each other. Photoelectrons (p.e.) can migrate from the hit pixel to another not directly fired by a photon. Optical cross-talk between pixels leads to a non-Poissonian behaviour of the distribution of fired pixels. 

An estimate of the optical cross talk probability can be obtained by the ratio double-to-single pulse rate as a function of the temperature. The probability depends weakly on the temperature and the measured level of cross-talk (15-16\%) is compatible with the one reported in the datasheet. 

\begin{figure}[t!]
 \centering
  \includegraphics[width=0.52\textwidth]{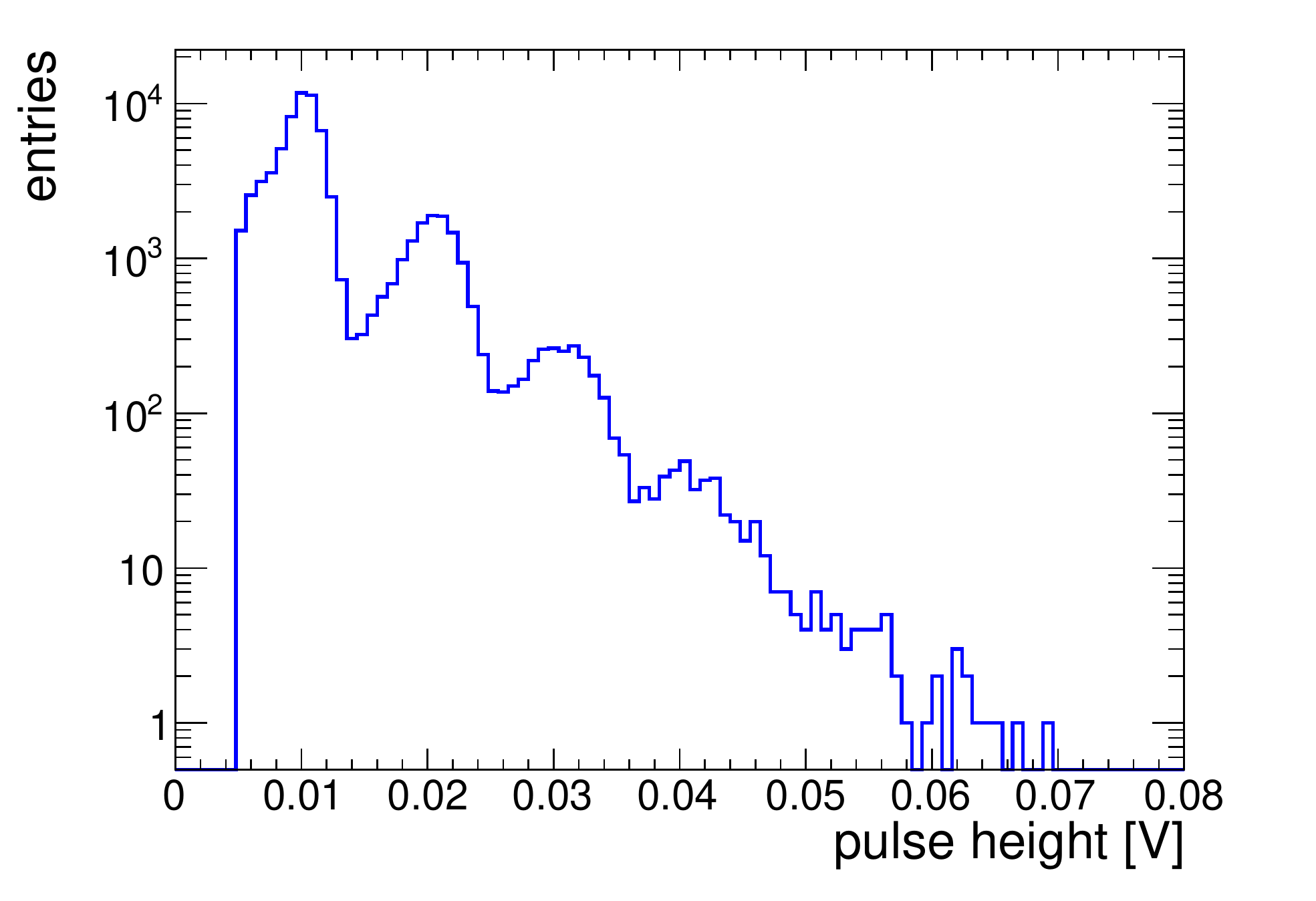}
 \caption{\footnotesize Dark current pulse height distribution obtained using GosSiP simulation package with the SiPM MicroSL-10035 parameters as input. } 
\label{fig:dark_simulation}
\end{figure}

The results obtained with the selected SiPM in different conditions have been compared with the output of the simulation framework GosSiP \cite{Eckert}, which gives a detailed model of the SiPM response once its basic parameters and cells configuration are given. In the Fig. \ref{fig:dark_simulation} it is shown the  pulse height distribution of the dark current for the SiPM under test.  

\section{Scintillating bar detector}\label{sec:scinti}
The scintillator bars considered for the design of the particle tracker are triangular in cross-section with a height of (17.0 $\pm$ 0.5) mm and width of (33.0 $\pm$ 0.5) mm, each with a (2.6 $\pm$ 0.2) mm diameter hole used to lodge a fiber to collect the light. The lateral surface of the scintillator strips is painted with white EJ-510 TiO$_{2}$ Eljen paint.

The scintillation light is collected with 1.2 mm BCF-91A WaveLength Shifter (WLS) fiber produced by the Saint-Gobain Ltd. \cite{saintgobain}. The WLS is glued into the hole running along the bar and its ends are polished. The read-out is performed by the SiPM only at one end and the opposite side is mirrored with reflecting tape to maximize the light collection.

\section{Front-end electronics}\label{sec:respsens}
The front-end board prototype dedicated to the amplification and SiPM readout has been developed by the Bologna INFN electronic group. The amplification system is a two-stages transimpedance amplifier which is used to convert the current coming from the SiPM in an output voltage: $V_{out} = - R*I_{SiPM}$. The current from the SiPM is discharged on the low input resistance of the transimpedance amplifier; this gives small time constants, that is, fast signal rise time (using the OPA 656N with a 500 MHz bandwidth we obtain signals with 20-30 ns of rise time). The transimpedance amplifier output is split in two chains: the first one is an integration and shaping chain with a time constant of about 150 ns and the second one is an amplification chain with gain G = 10. 

The final printed board handles 8 channels, giving a very clean output both for linear and integrated pulses. The amplitude of the electronic noise fluctuations is less than 10\% of pulse heights of a single pixel. 

\section{Study of the detector response and sensitivity}
The response of each bar coupled with a SiPM has been tested by injecting light pulses provided by a laser and also by collecting light induced by a radioactive source.

\subsection{Tests with laser pulses}\label{sec:laser}
The product of the photon detection efficiency (PDE) and the gain determines the SiPM response to the light hitting its surface. For the SiPM under test the PDE, which is wavelength dependent, is about 15\% for $\lambda \sim$ 500 nm and the gain can be calibrated precisely by evaluating the distance between two adjacent peaks of a charge spectrum, corresponding to the charge of 1 pixel fired.

The gain depends on V$_{bias}$ and on the temperature via the breakdown voltage. At our working conditions (V$_{wk}$ = 29 V and room temperature T $\sim$ 26 $^{\circ}$C) the gain is $\sim$ 6 $\times$ 10$^{6}$ and the corresponding amplitude of a pixel signal is $\sim$ 10 mV. We have measured also the variation of the peak amplitude dependence on T which is about 0.25 mV/$^{\circ}$C.

\begin{figure}[t!]
 \centering
  \includegraphics[width=0.48\textwidth]{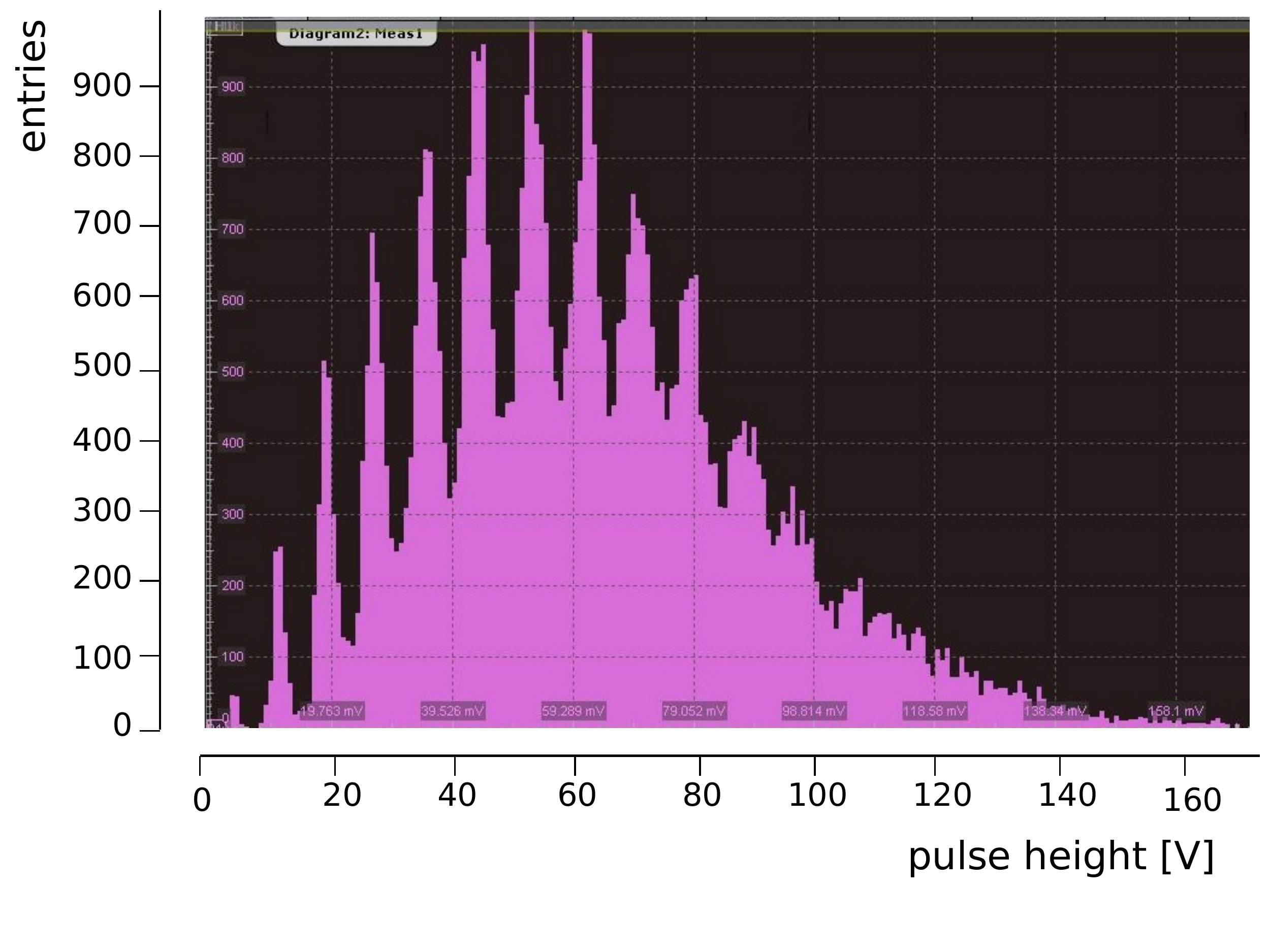}
 \caption{\footnotesize Charge spectrum of a SiPM obtained injecting laser pulses on the scintillator bar at a distance of $\sim$ 6 mm from the WLS fiber. Signal amplitude peak corresponds to 6 pixels while the individual pixel value is $\sim$ 10 mV. }
\label{fig:laser_calibration}
\end{figure}

Firing a SiPM by an attenuated spot of the Picosecond Diode Laser (PiLas) 407 nm \cite{PiLas} we obtained the charge spectrum shown in Fig. \ref{fig:laser_calibration}.

\begin{figure}[t!]
 \centering
  \includegraphics[width=0.52\textwidth]{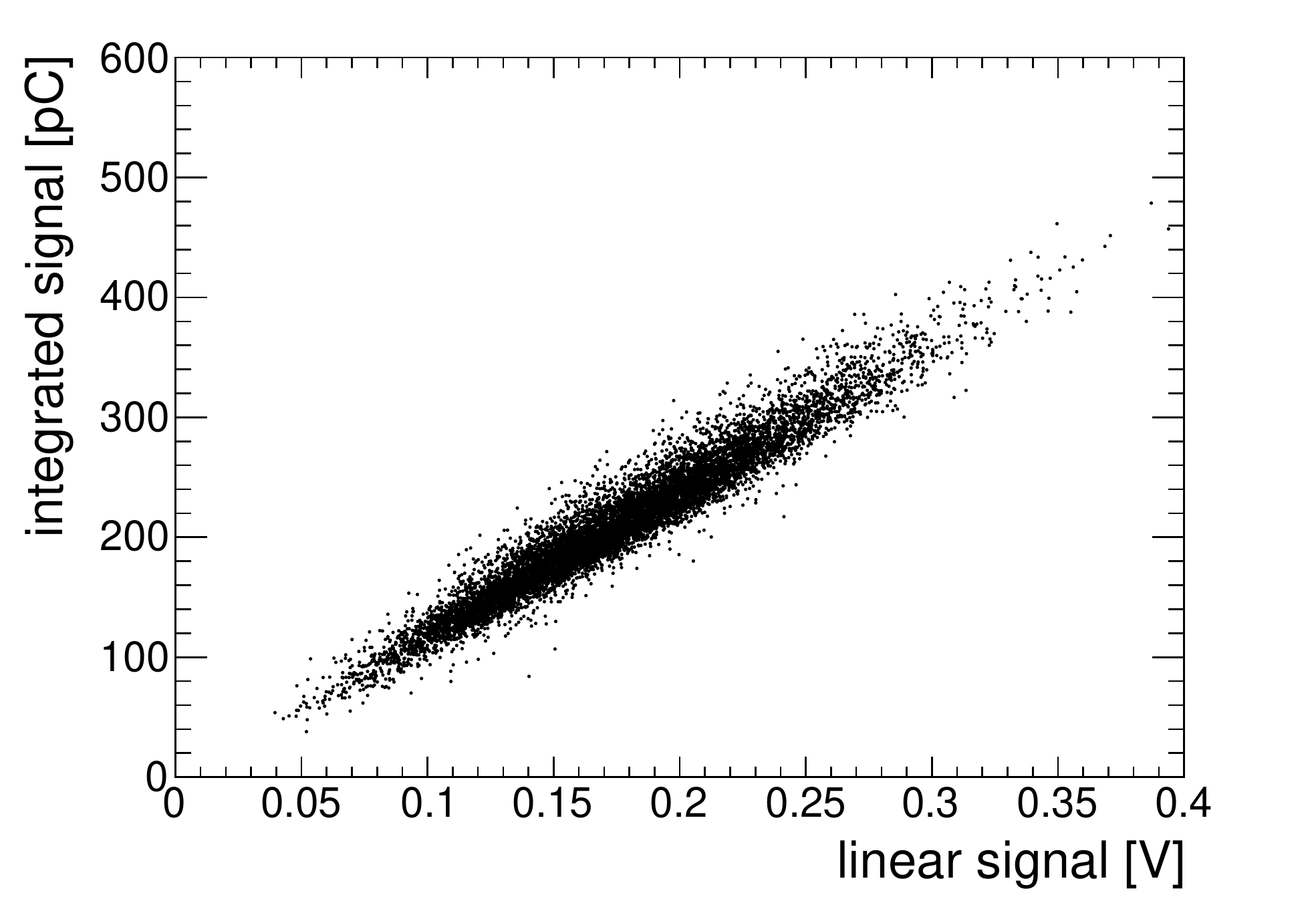}
 \caption{\footnotesize Integrated signal value vs. peak amplitude. }
\label{fig:lin-int_signals_dig}
\end{figure}

In our tests we deal with Minimum Ionizing Particles (MIP) which produce in the scintillator a small flux of light signals \cite{montanari}. Thus, no problems are expected concerning the dynamic range and linearity.

We note here that we obtain a precise proportionality between the height of voltage pulse and its integral as shown in Fig. \ref{fig:lin-int_signals_dig}, so we will use indifferently mV or ADC counts to indicate the SiPM outputs.

First, all bars were tested with laser pulses. The SiPM response was checked and WLS-SiPM coupling investigated as potential source of systematic loss of signal. Intercalibration factors for the bars were obtained ranging from 5 to 10 \%. These factors are used to correct raw data in the analysis.

A MC simulation has been performed based on the GosSiP package \cite{Eckert} in order to compare the predicted signal amplitude distribution to the observed ones of Fig. \ref{fig:laser_calibration}. A good agreement between the two distributions has been found when simulating a single laser pulse composed by a fixed number of photons impinging uniformly on the SiPM sensitive surface. 

\subsection{Tests with a radioactive source}
 We studied the response of a system made of two adjacent triangular bars (see Fig. \ref{fig:reco_pos_princ}) coupled to SiPMs using a $^{90}$Sr $\beta-$source having an activity of $\sim$ 90 kBq. The two 50 cm long scintillators were placed inside a metallic box with 5 equispaced holes for accommodating the source at different points along the two bars. The vertical through the collimated source passed at a distance of 4 mm from the fiber of scintillator 2 and 12.5 mm from that of scintillator 1 (see Fig. \ref{fig:reco_pos_princ}). The SiPMs were cooled using Peltier cells and, in order to have the best signal-to-noise ratio, we chosed a temperature of 1 $^{\circ}$C. With this setup a signal is acquired triggering on the coincidence in both scintillators.

\begin{figure}[t]
\centering
\subfloat[][\emph{}]
{\includegraphics[width=.42\textwidth]{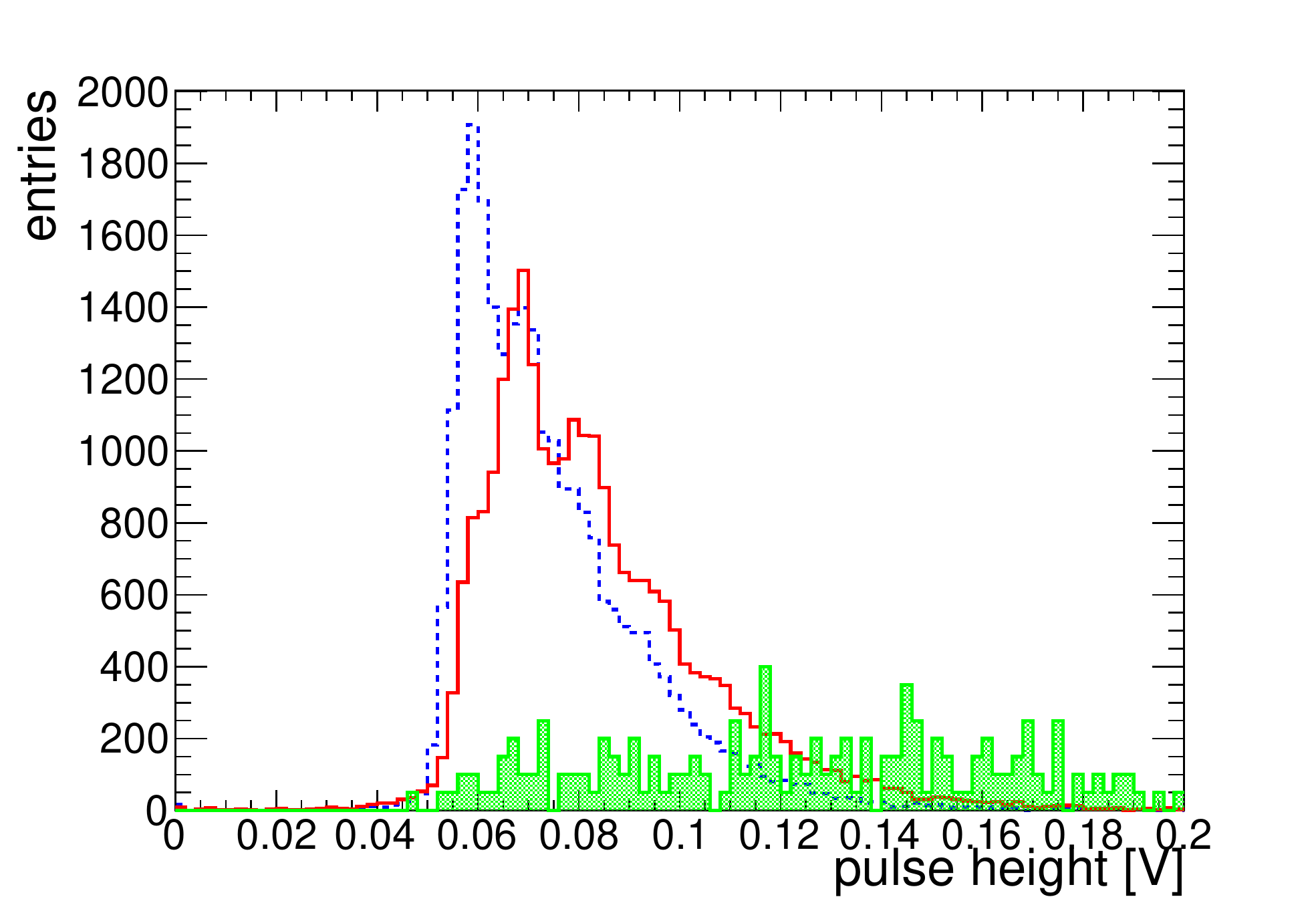}} \quad
\subfloat[][\emph{}]
{\includegraphics[width=.40\textwidth]{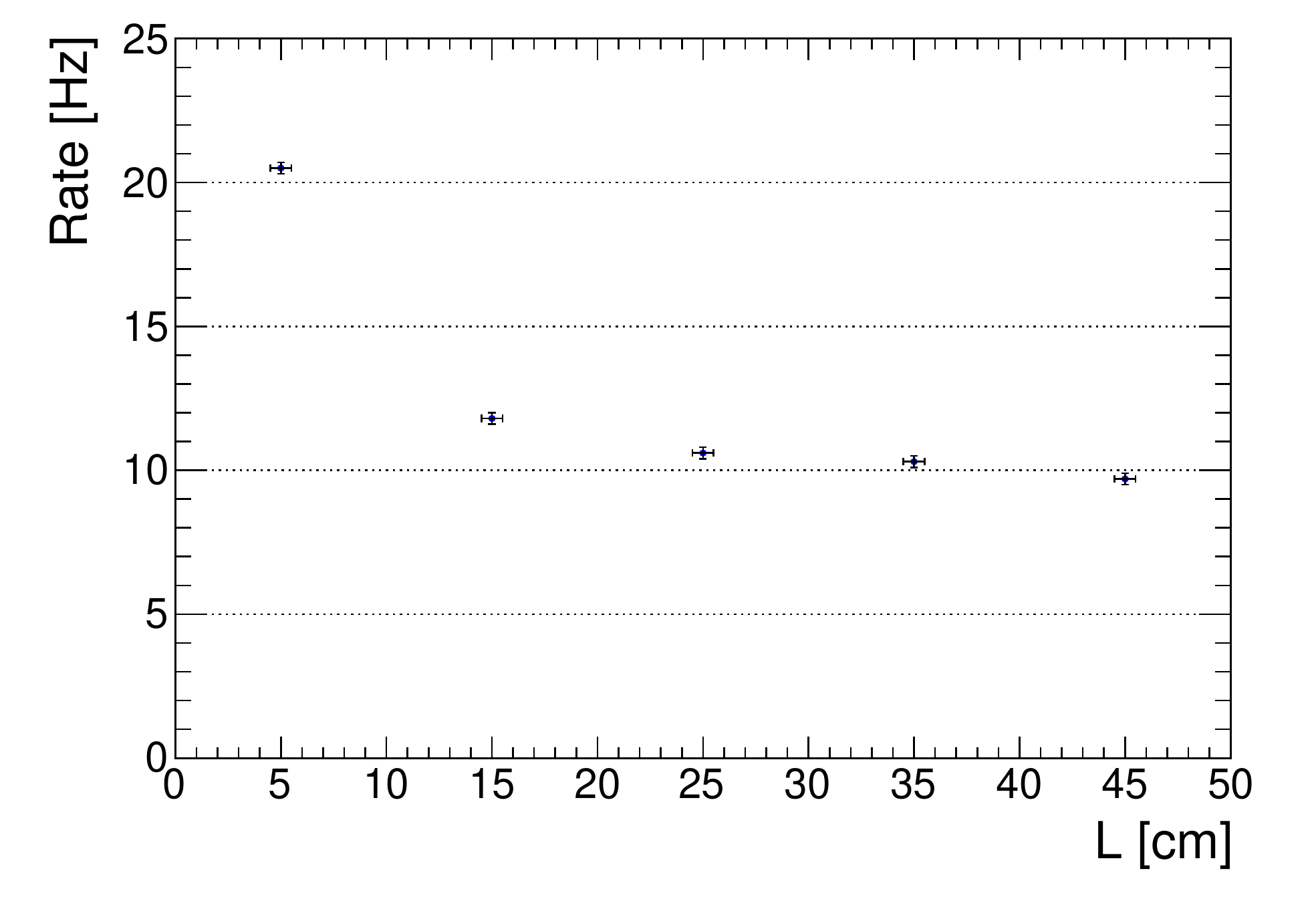}}
\caption{\footnotesize (a) The distributions of pulse height measured with a radioactive source placed over the two bars (see text) at a distance of 25 cm from the SiPMs: red solid (blue dashed) curve for scintillator 1 (2); green filled area corresponds to the coincidences of scintillator 1 and 2 observed without the source. The latter is the background due to CRs particles. In this case the values have been multiplied by a factor 50 to make it evident and its contribution is clearly negligible. (b) Rates of high pulse (V$_{th} $ = 80 mV) coincidences of scintillator 1 and 2 with the source at different distances from the SiPM.}
\label{fig:subfig_source}
\end{figure}

In Fig. \ref{fig:subfig_source}a is shown the distribution of pulse height signals obtained with and without radioactive source.

As the reduction in the pulse peak amplitude is proportional to the light attenuation we evaluate the ``attenuation" length of our bars by measuring the coincidences rate of pulses over a certain threshold when the radioactive source is placed at different positions along the two scintillator bars arranged as before. The rate variation as function of distances from the SiPM is shown in Fig. \ref{fig:subfig_source}b.

By taking into account a systematic uncertainty of 1 \% on the longitudinal positioning of the source along the bar and of 2 \% due to the CRs contribution, the fit to the data of the sum of two exponential functions gives a ``short" (4.4 $\pm$ 0.5) cm and ``long" (290 $\pm$ 70) cm ``attenuation" lengths, the former being probably dominated by self-absorption. We stress that our bars are only 50 cm long so it is difficult to make a direct comparison with other results. 

\begin{figure}[t!]
 \centering
  \includegraphics[width=0.35\textwidth]{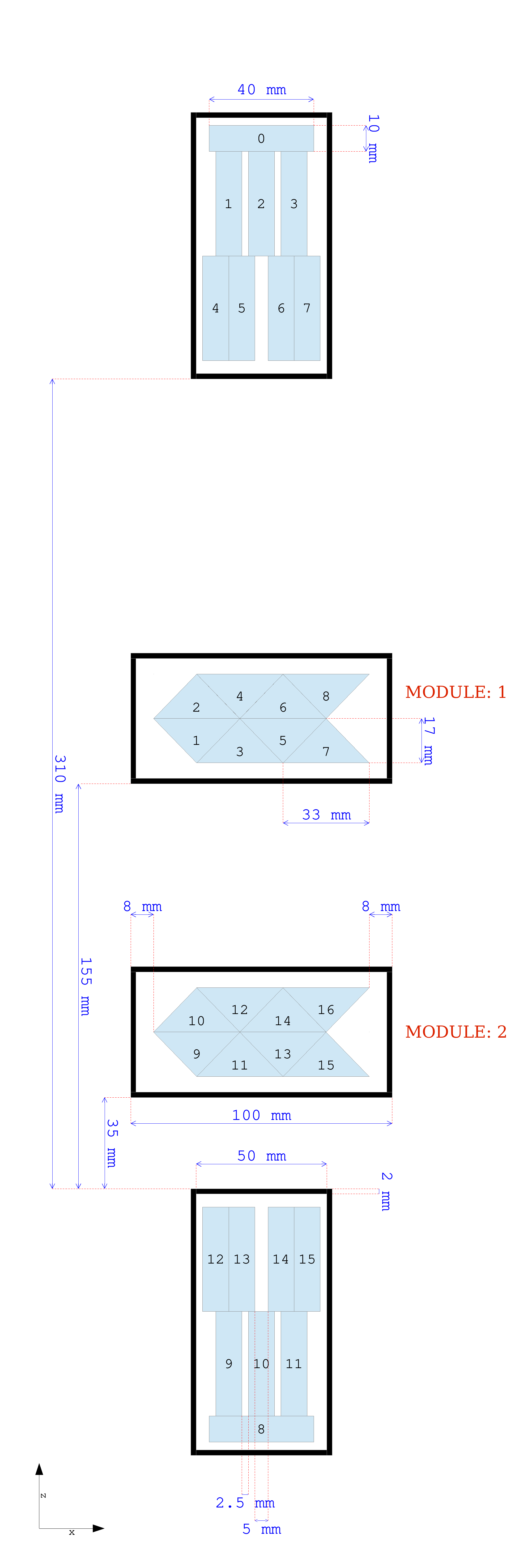}
 \caption{\footnotesize The set-up of scintillating bars (rectangular for the trigger and triangular for particle trajectory reconstruction) used for detecting CR particle. The  (upper and lower) triangular planes can be independently moved in the x-direction of a fraction of a mm. The triangular plane in module 1 and 2 are identified as P1, P2, P3 and P4 starting from top plane. The staggered configuration of the rectangular bars (60 $\times$ 4 $\times$ 1 cm$^{3}$) allows to select CR particles in a window of 0.5 and 0.25 cm. The length of triangular bars is 50 cm. We set the origin of axis in the $x$ direction at the external border of the aluminum box of module 1 and 2.}
\label{fig:station}
\end{figure}

\section{Tests with Cosmic Rays}

\subsection{Layout}
 In order to study the spatial resolution achievable with a tracking device composed of planes of triangular bars equipped by SiPMs in analog readout mode (described in section \ref{sec:scinti}) a simple tracking system has been set up. It consists of two modules with a distance of about 12 cm, each composed of two faced planes of 4 scintillator bars (50 cm long) (see Fig. \ref{fig:station}). The modules are inserted in an aluminum box closed by a cap with plugged in SiPM in order to guarantee a good coupling between the sensor and the fiber. The 8 channels front-end board is soldered on the external edge of the cap. Triangular bars in each module are tied together and taken in place by the box and mask. 

To select CR particles crossing the detector in well defined positions and in almost vertical directions, an external trigger system was added. Tracking planes are located between two trigger stations separated by $\sim$ 40 cm. Each trigger station is composed of 8 staggered scintillator rectangular bars (60 $\times$ 4 $\times$ 1 cm$^{3}$ each) equipped with wavelength-shifting fibers and SiPMs provided by AdvanSiD (ASD-SiPM1S-M-100) \cite{adv}. Scintillator bars, SiPM, readout and amplification board are provided by the INFN-LNGS electronic group. As can be seen in Fig. \ref{fig:station} there are 6 possible quadruple trigger configurations that can select tracks in windows 5 mm or 2.5 mm wide.

The two modules of triangular bars could be displaced horizontally with respect to the fixed rectangular bars by operating a fraction of mm precision rod screw. The overall uncertainty in nominal up/down relative positions of triangular bars was estimated less than 1 mm.

\subsection{DAQ}
 A simple DAQ system has been used based on NIM and VME standards. All rectangular bars signals are sent to discriminators and put in coincidence to select well defined tracks: the coincidence signal is then used as a trigger to acquire triangular scintillators signals by means of waveform digitizers. The CAEN 12 bit 250 MS/s mod V 1720 Digitizer was used thus allowing a time window of 4 $\mu$s with 4 ns sampling time. This module has also an internal generating self-trigger capability and acquisitions were done also excluding the external rectangular trigger system (autotrigger mode). 

The triangular and rectangular channels are also recorded by pattern units and scalers. A standard interface VME-PC permits run controls and the storage of the acquired data.

\subsection{Position reconstruction}
\begin{figure}[t!]
 \centering
  \includegraphics[width=0.40\textwidth]{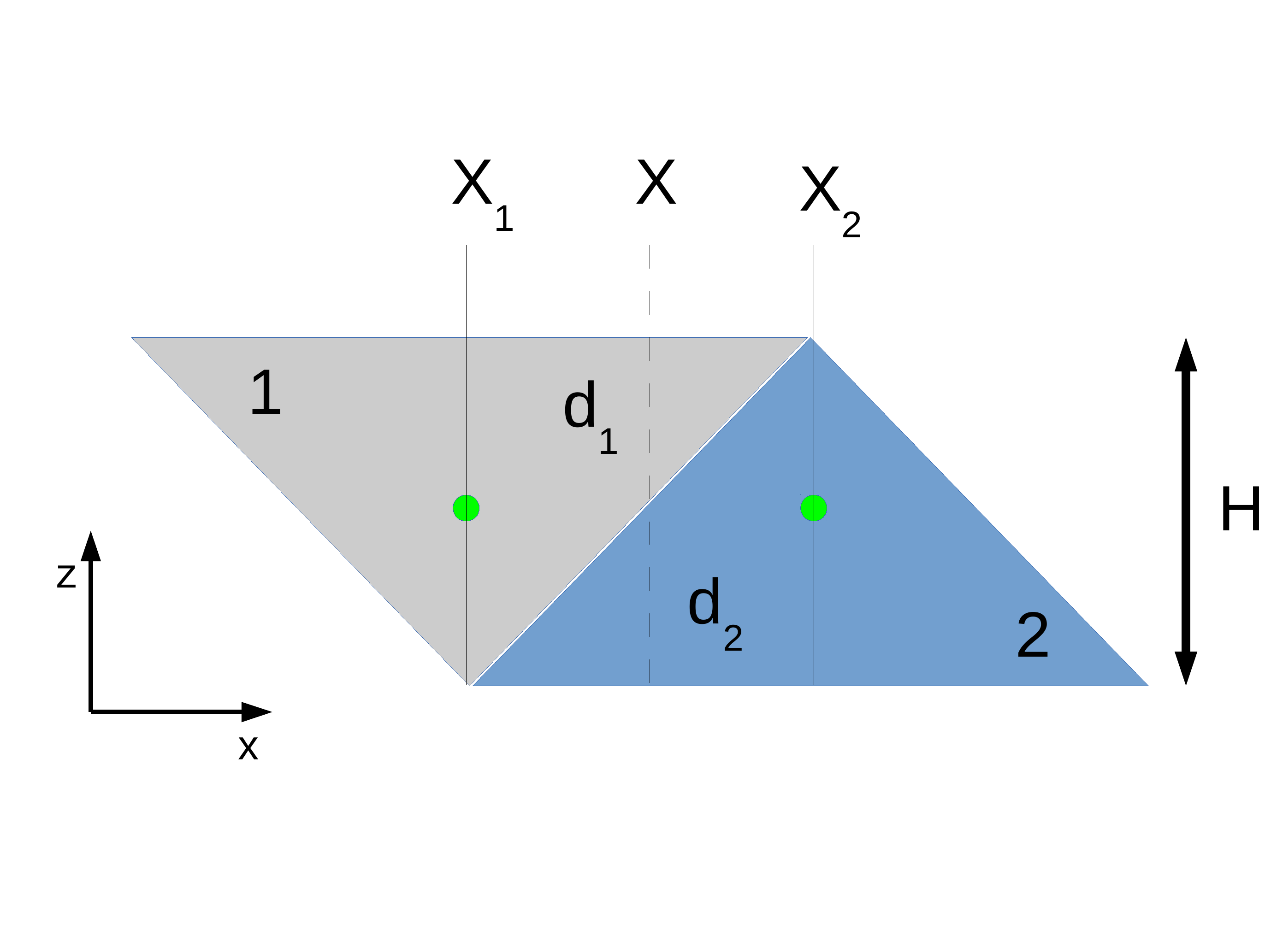}
 \caption{\footnotesize Scheme used for reconstructing the X-position of the crossing particle. X$_{1}$ and X$_{2}$ are nominal positions of two adjacent triangular bars (in the case shown the coordinates of the WLS).}
\label{fig:reco_pos_princ}
\end{figure}

 The particle crossing point on a plane of triangular bars is obtained by reading the pulse height in each channel. The principle for the algorithm used is shown in Fig. \ref{fig:reco_pos_princ} and is expressed by the following formula:
\begin{equation}\label{eq:reco_pos}
X_{rec} = (E_{1}X_{1}+E_{2}X_{2})/(E_{1}+E_{2})
\end{equation}
where $E_{i}$ is the energy lost by the particle crossing the scintillator bars; $X_{i}$ is the nominal position of the WLS in each triangular bar from a reference. With the assumption that the pulse height $w_{i}$ is proportional to the path length of the particle across the bar ($w_{i} \propto d_{i} \propto E_{i}$) and $d_{1}+d_{2}$ = H (see Fig. \ref{fig:reco_pos_princ}), the position of the crossing point in the plane of the fibers is calculated as:
\begin{equation}\label{eq:reco_pos2}
X_{rec} = (w_{1}X_{1}+w_{2}X_{2})/(w_{1}+w_{2})
\end{equation}

We will use the relation (\ref{eq:reco_pos2}) also for track reconstruction of particles having momentum components in the $y$ direction, as the relative path length of the selected crossing particle in the bars is the same.

The coefficient of proportionality between the pulse height and the energy deposited in the bar ($w_{i} = c_{i}E_{i}$) can vary for each bar (e.g. WLS coupling with SiPM, scintillator response, etc), thus affecting the position reconstruction. The equalization coefficients must be evaluated in order to obtain uniformity in pulses height between adjacent bars.

\subsection{ Data Taking - External trigger mode}
\subsubsection{Run conditions}
 The single counting rates of each triangular bar (for signal above a threshold of about 3 pixels) have been monitored over long periods of time in order to study the stability of the system. Quite stable behaviours for these rates were obtained; night/day effects were observed due to slightly variations on room temperature.

The trigger on rectangular bars was defined by a quadruple coincidence of vertical staggered scintillators (see Fig. \ref{fig:station}). A conservative threshold was set to ensure the cut of any noise signal. A rate of $\sim$ 20 events per hour was collected by 5 mm window and $\sim$ 1 event per hour by 2.5 mm window.

\begin{figure}[t!]
 \centering
  \includegraphics[width=0.52\textwidth]{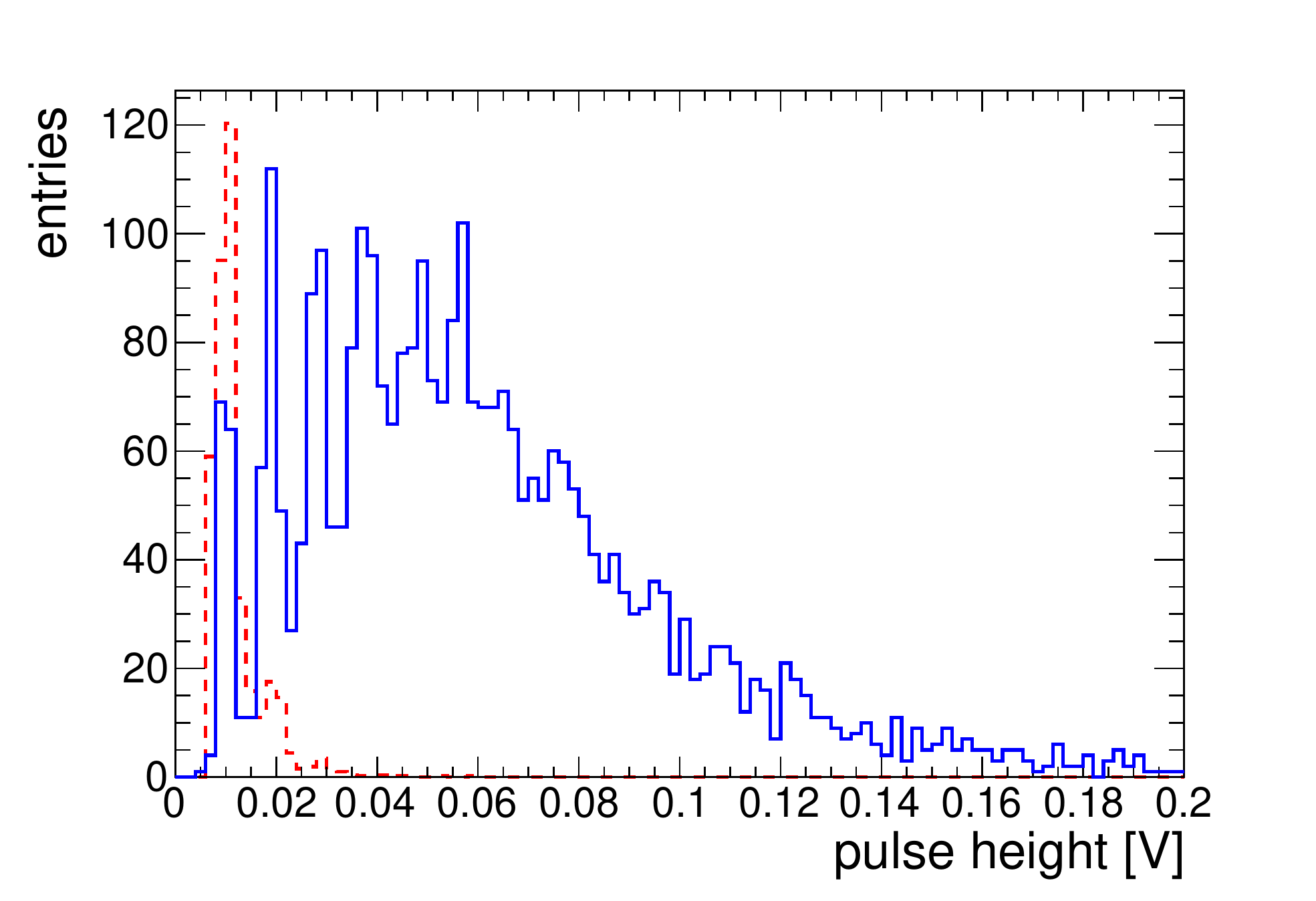}
 \caption{\footnotesize Signal (blue solid histogram) to noise (red dashed histogram) comparison. The latter distribution contains signals collected in a time interval of duration identical to that of CR events, far from the trigger time window.}
\label{fig:final}
\end{figure}

The output signals of the triangular bars on time with CR particles selected by the trigger system have been compared to the dark current signal amplitude in Fig. \ref{fig:final}. We accept as on time the signals reaching their maximum before 10 digitizer sample counts (40 ns) from trigger time. The distribution of the signals generated by the passage of particles in the triangular bars appears well separated by the noise. 

At the work conditions fixed for the SiPMs described in section \ref{sec:SiPM}, gain for 1 pixel leads to $\sim$ 10 mV amplitude. The use of a Peltier cell can easily reduce the noise by a factor of 5 (from 900 kHz to about 200 kHz). Output signals of our bars have a rise time of about 20-30 ns. At a temperature of $\sim$ 26 $^\circ$C the rate of 1 dark pixel represents a few \% of all collected signals. So in these limits and for our aims we have chosen to run at a normal room temperature. 

\subsubsection{Position reconstruction}
 To evaluate the reconstructed position of tracks and the spatial resolution achievable with triangular scintillators bars we made a series of measurements introducing an external trigger generated using the upper and lower rectangular scintillator bars. First the two modules of triangular scintillator were aligned with respect to the trigger system (see Fig. \ref{fig:station}). Then various displacements were chosen in order to select tracks of CR muons crossing different pairs of adjacent triangular scintillator bars. 

As an example, in a run of about 560 hours we collected 3568 events triggered by quadruple 3-6-11-14 (see Fig. \ref{fig:station}) requiring via software the absence of any other rectangular bar fired. This ensures a sample of CR events selected in a 5 mm wide window. In this configuration modules 1 and 2 are shifted with respect to the external trigger station in such a way to select tracks that cross the triangular channels 1, 2, 3, 4  and 9, 10, 11, 12 (see Fig. \ref{fig:station}).

\begin{figure}[t!]
 \centering
  \includegraphics[width=0.52\textwidth]{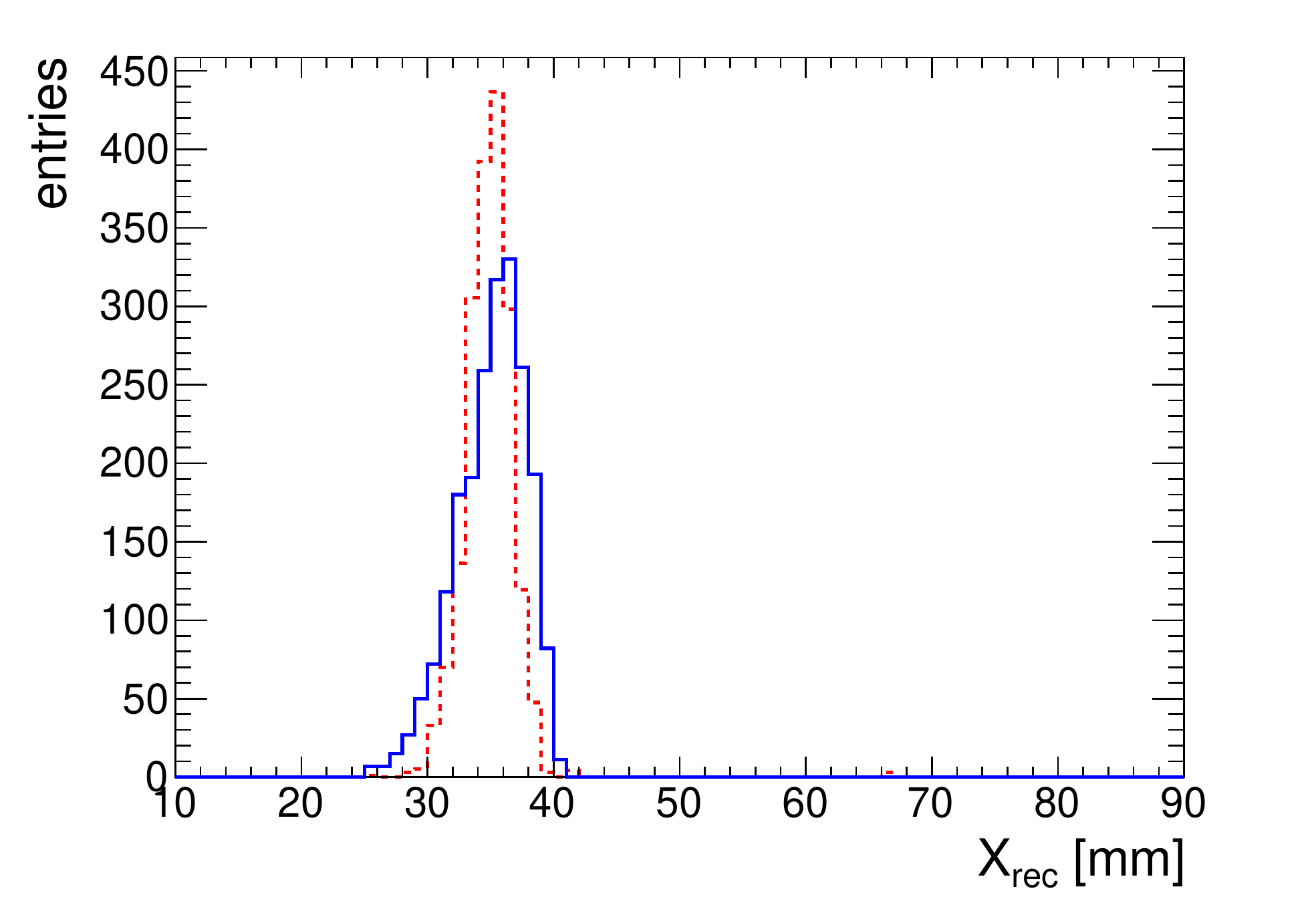}
 \caption{\footnotesize Distributions of reconstructed positions in the P2 plane (see Fig. \ref{fig:station}) for events selected by the quadruple trigger 3-6-11-14 and with only two adjacent triangular bars fired: data (blue solid) and MC (red dashed) histograms. The FWHM of the data distribution results 5.5 mm to be compared to the value of 4 mm predicted by MC simulation. $X_{rec}$ is referred to the external border of the aluminum box of module 1 and 2 (see Fig. \ref{fig:station}).}
\label{fig:subfig_pos_361114}
\end{figure}

For those events we calculated the reconstructed position of a track in each single plane of triangular bars for the subsample of events in which two adjacent triangular bars have a signal on time with the external trigger. The distribution of reconstructed positions in the P2 plane of the up module using only such events is shown in Fig. \ref{fig:subfig_pos_361114}.

A MC simulation has been set-up to understand the behaviour of CR in the geometry of the apparatus. Both trigger and tracking system are taken into account as described in the previous section.

By using Geant 4.9.5 \cite{geant} muons crossing the system and having the following characteristics have been simulated:
\begin{itemize}
\setlength\itemsep{0em}
\item E$_{\mu} =$  1 GeV;
\item Position distribution uniform on the triangular scintillators surface;
\item Azimuth direction uniformly distributed;
\item Zenithal distribution like cos$^{2}\theta$;
\end{itemize}

This MC simulation takes into account only geometrical effects and energy losses. Only the hits due to muons are considered and no threshold on the deposited energy is used. In Fig. \ref{fig:subfig_pos_361114} the experimental distribution has a FWHM of 5.5 mm to be compared with the MC prediction of 4 mm.

\begin{figure}[t!]
 \centering
  \includegraphics[width=0.52\textwidth]{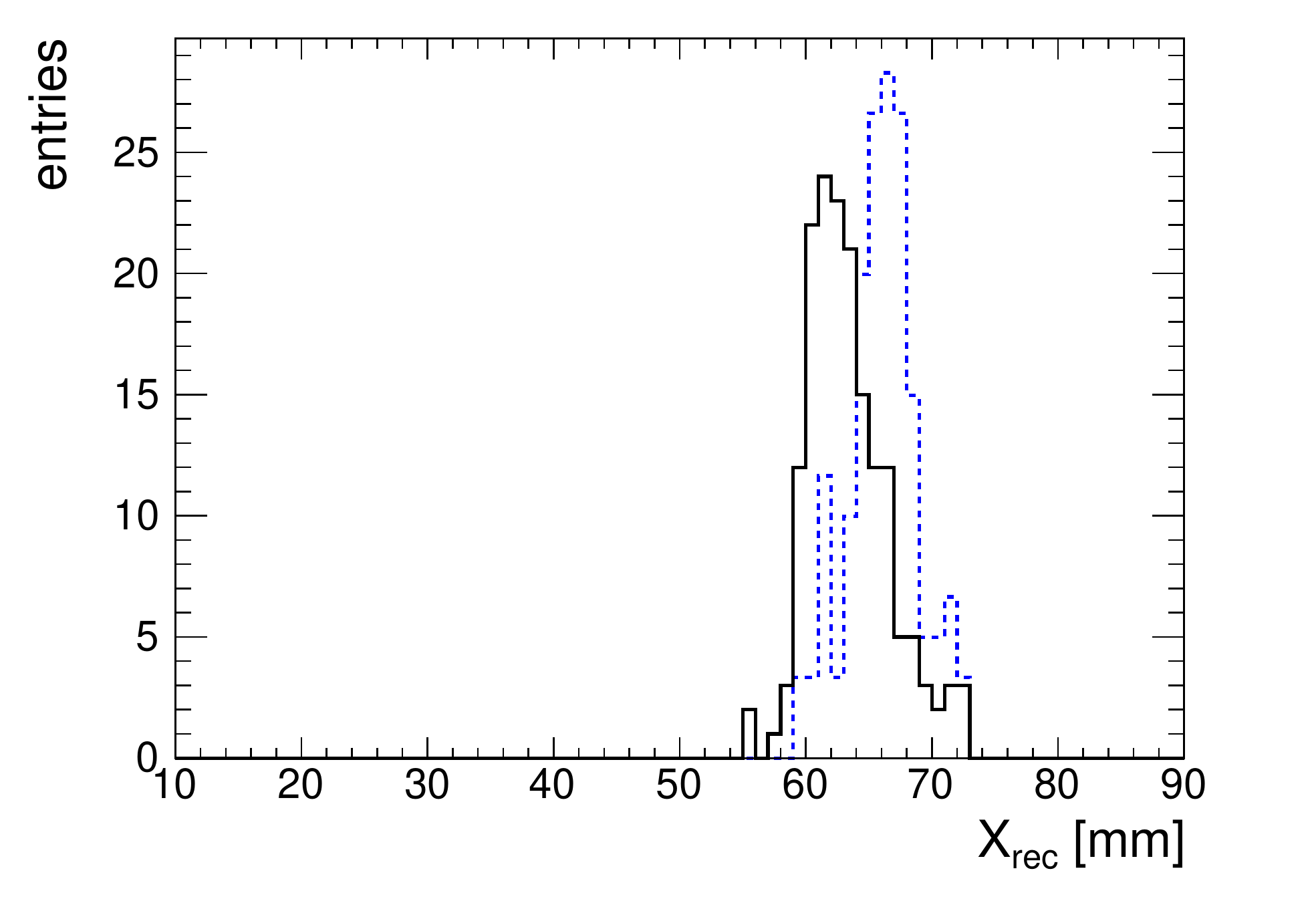}
 \caption{\footnotesize (black solid histrogram) distribution of reconstructed position in the P2 triangular bar plane for events selected by the quadruple trigger 2-6-10-14 and with only two adjacent scintillators fired; (blue dashed histogram) the distribution obtained after shifting the module 1 by 3 mm to the left with respect to the fixed external trigger system. The FWHM of the former is 6 mm and of the latter is 4.5 mm while the distance between the barycenter of the two distribution is 2.7 mm.}
\label{fig:subfig_pos_261014}
\end{figure}

We took also two runs with two different configurations of the triangular bar modules with respect to the external trigger frame. In the first run the left walls of the triangular bar module boxes were aligned with the left sides of the rectangular bar boxes. We select events triggered by the quadruple 2-6-10-14 (2.5 mm wide window) and signals from channels 1, 2, 3, 4 were collected. Then we shifted to the left by 3 mm the module 1 and we collected signals from channels 5, 6, 7 and 8 (see Fig. \ref{fig:station}). The distributions of reconstructed event positions in the aforementioned module alignments are compared in Fig. \ref{fig:subfig_pos_261014}. The result demonstrates the capability of the triangular bar system in resolving particle tracks.

\subsection{ Data Taking - Autotrigger mode}
\subsubsection{Trigger with triangular bars}
 We have also investigated the angular resolution achievable by triggering with the 4 planes of module 1 and 2 only.
The following trigger settings have been used:
\begin{itemize}
\setlength\itemsep{0em}
\item a threshold on each channel was set to about 5 pixel (50-55 mV)
\item at least one triangular bar for each plane over the threshold
\item all planes have to be fired
\end{itemize}

 The threshold was fine tuned in order to have almost the same single signal rate in each channel and to ensure that if a track crosses a couple of adjacent bars, at least one bar has a signal over the threshold.

\begin{figure}[t!]
 \centering
  \includegraphics[width=0.42\textwidth]{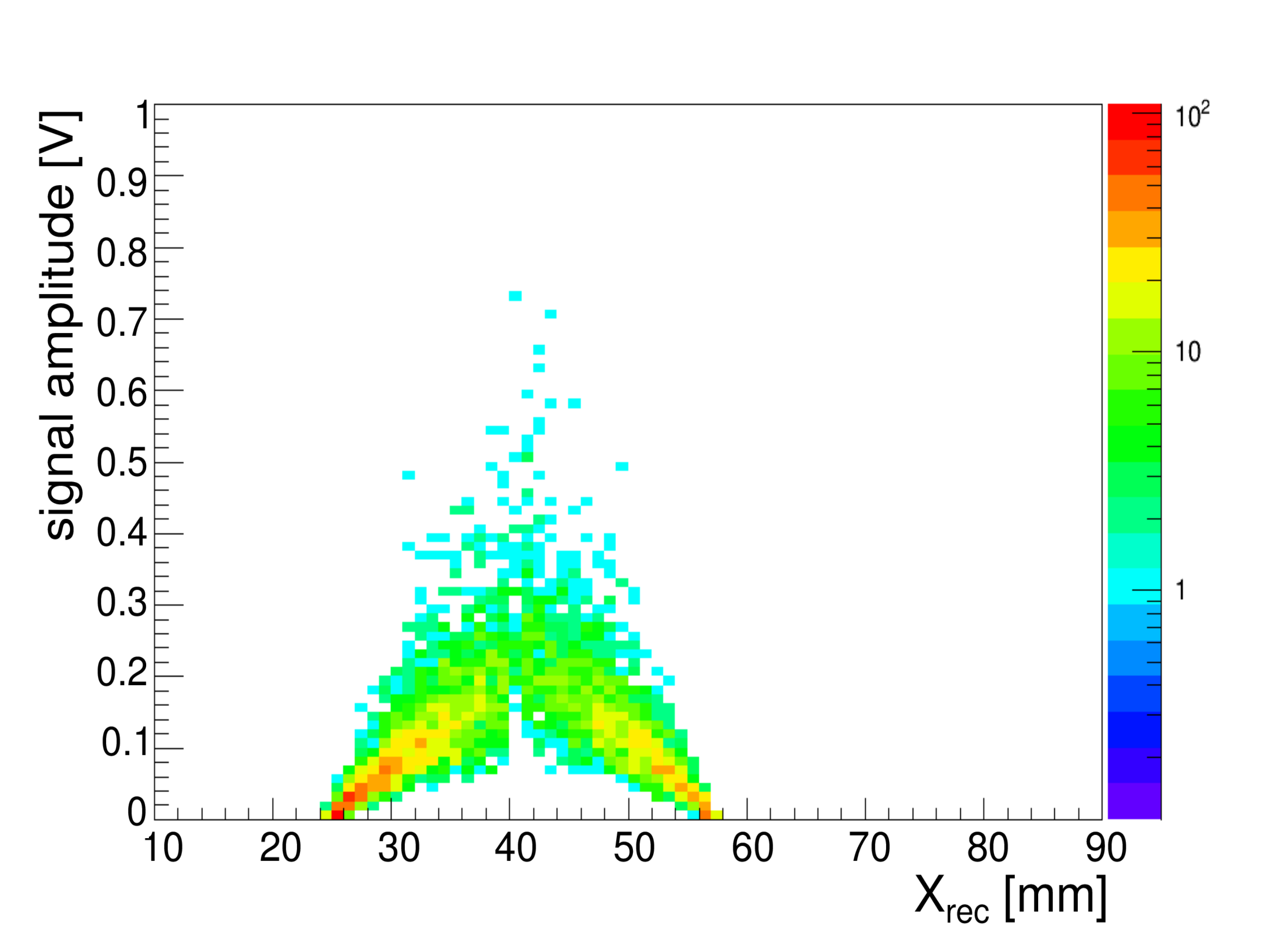}
 \caption{\footnotesize Signal amplitude vs. reconstructed position in the plane P2 for events with 2 adjacent bars fired (1-3 and 3-5) in that plane. No cut on signal amplitude was applied. }
\label{fig:signal_posRecp}
\end{figure}

The signal amplitude in the triangular channel 3 as function of the reconstructed positions in the plane P2 for events crossing the pairs of adjacent triangular bars 1-3 and 3-5 is shown in Fig. \ref{fig:signal_posRecp}. Clearly a cut on the signals would enlarge the ``dead zone" at the vertexes of the triangular bars, namely region where it is not possible to assign the ``true" position to a track, but the nominal of the fiber.  For example, if a cut of 3 pixel is applied, a dead zone of about 1.5 mm will be present around the fiber nominal position.

However in our analysis we have further applied an offline cut on signal amplitudes by selecting pulses $>$ 15 mV in all bars. In this way the reconstructed position is not affected by the contribution of 1-pixel spurious dark pulses that randomly occurs in bars not participating in the trigger.

\begin{figure}[t!]
 \centering
  \includegraphics[width=0.48\textwidth]{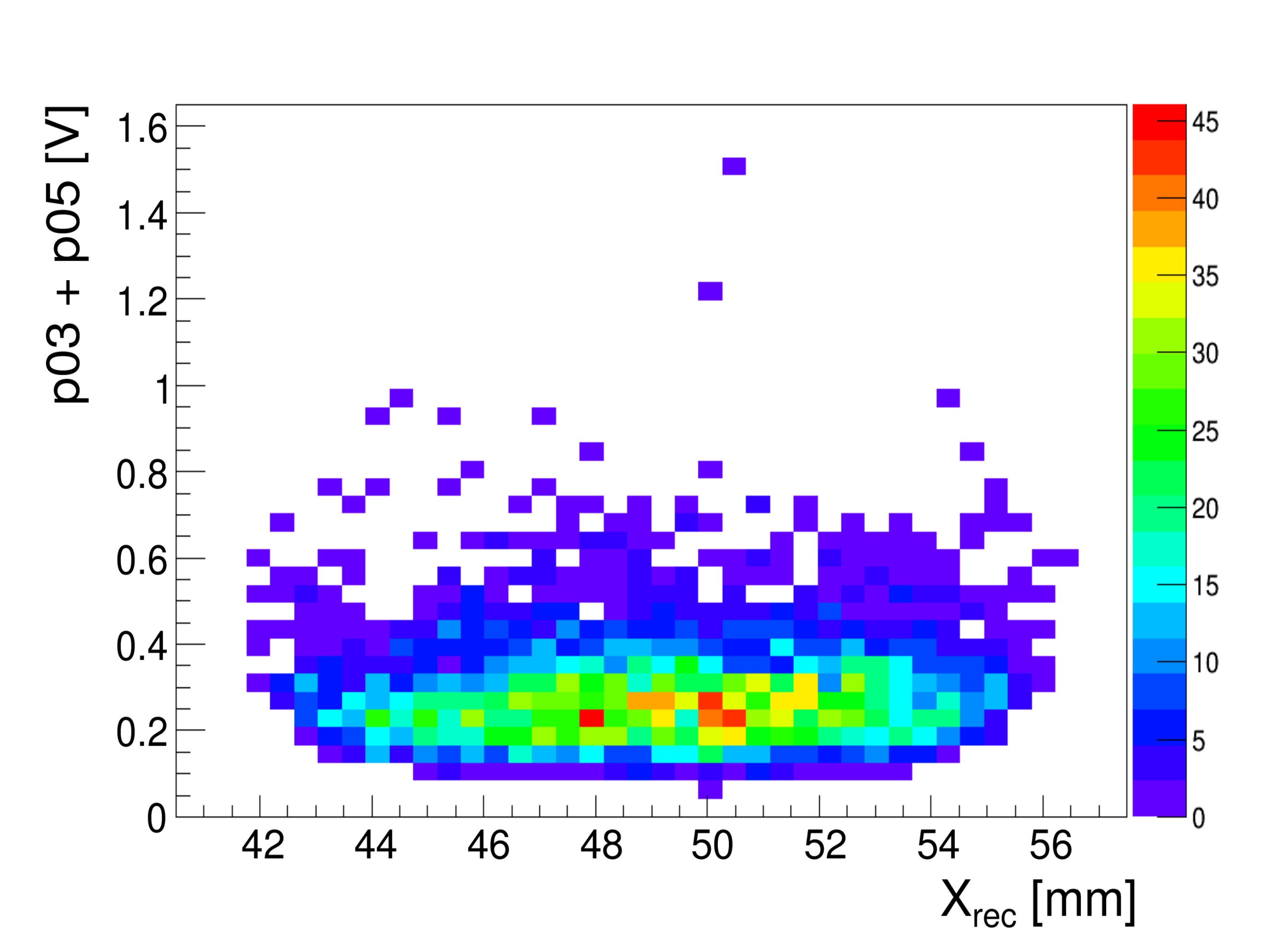}
 \caption{\footnotesize Sum of pulse heights in couple of adjacent triangular bars (3-5, 5-7) vs. the position reconstructed in the plane P2. Events with only two adjacent triangular bars in each plane and with a slope less than 50 mrad are selected. }
\label{fig:sum_pulse_posreco}
\end{figure}

The energy deposited by a vertical 1 GeV muon crossing an extruded scintillator bar is about 1.7 MeV/cm and the light yield is assumed to be 66\% of the BC408 scintillator \cite{fermilabconf}. Thus about 8500 photons/MeV are produced. A conservative evaluation of fiber BCF-91A light collection can be estimated to 0.4 \% \cite{aida}. This factor leads to about 60 photons/cm reaching the SiPM surface and considering a PDE of about 15 \%, a signal of a m.i.p. crossing 1 cm would correspond to about 9 p.e., i.e. 9 pixels fired. Adding the amplitudes of signals from two adjacent bars is almost equivalent to consider vertical tracks of particles crossing 1.7 cm of scintillator. Fig. \ref{fig:sum_pulse_posreco} shows the distribution of sum of pulse heights vs reconstructed position in a plane. The mean signal results to be 200 mV which corresponds to 20 p.e. detected in agreement with what expected.

\begin{figure}[t]
\centering
\subfloat[][\emph{}]
{\includegraphics[width=.42\textwidth]{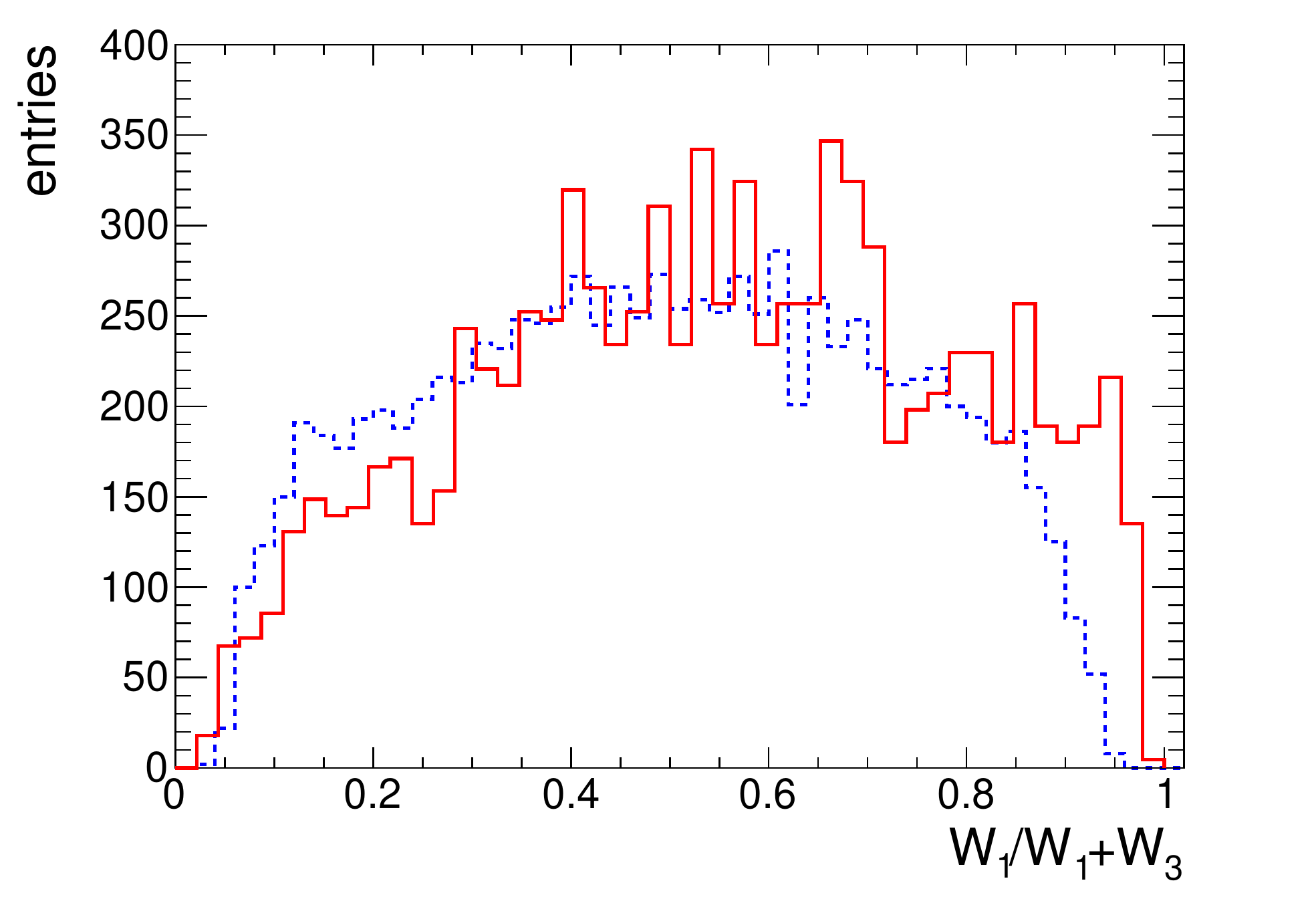}} \quad
\subfloat[][\emph{}]
{\includegraphics[width=.42\textwidth]{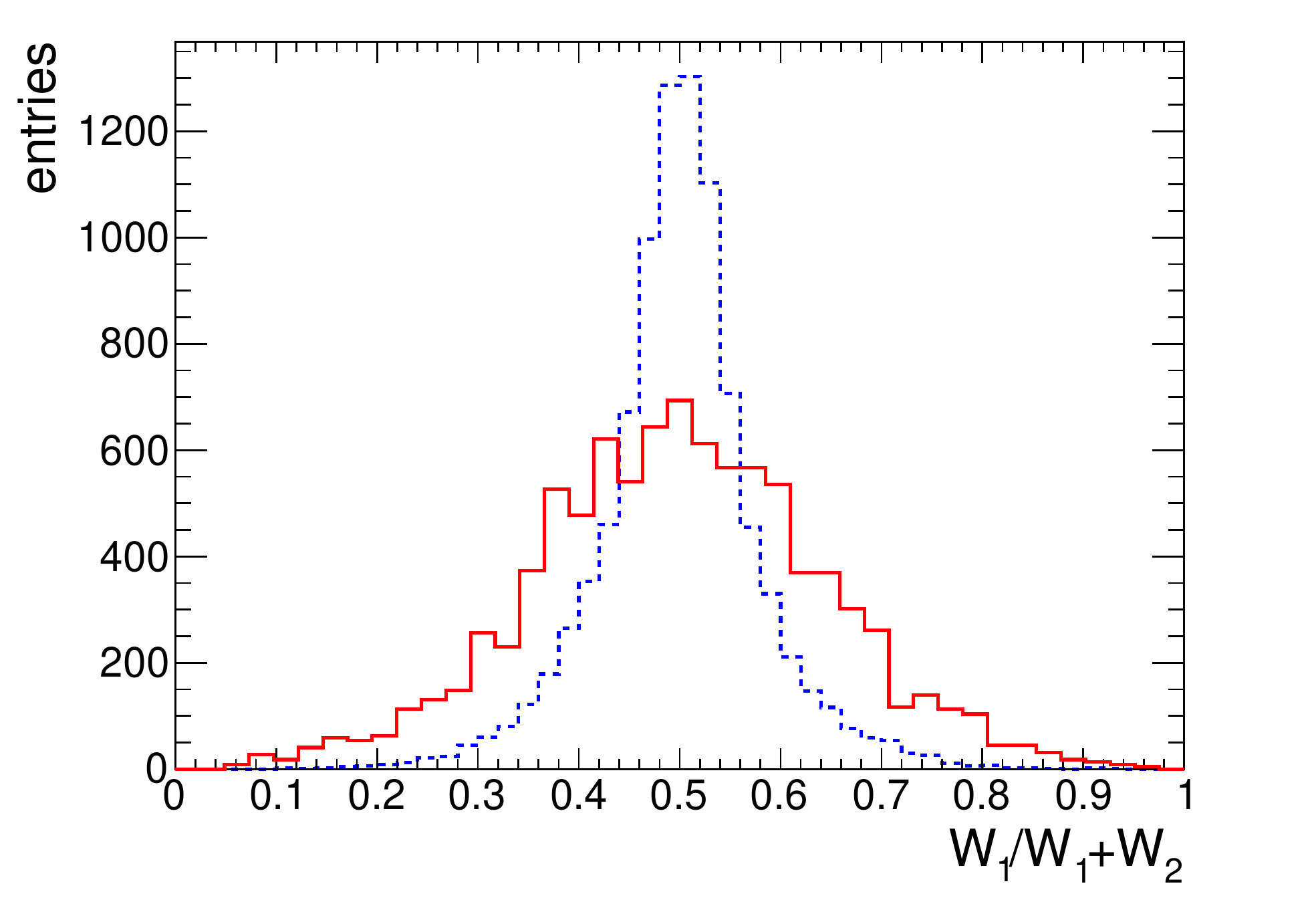}}
\caption{\footnotesize Comparison of the estimator $Q_{A} = W_{1}/(W_{1}+W_{3})$ (figure (a)) and $Q_{B} = W_{1}/(W_{1}+W_{2})$ (figure (b)) from data (red solid line) and MC (blue dashed line). }
\label{fig:subfig2}
\end{figure}

\subsubsection{Bar response equalization by data}
 A total of about 261000 events were collected at a rate of 1.7 Hz in autotrigger mode. In the following analysis a subsample was used, i.e. the events with only two adjacent triangular bars fired in each plane. This selection reduces the events to  about 22300.

In each module of triangular bars faced planes have not been staggered. This choice permits to cross check on the data the quality of signal distributions and the accuracy of our calibration system.

As described in section \ref{sec:laser} our laser calibration factors do not take into account all systematic uncertainties of the energy collection by the SiPM. So profiting from the couple of faced planes and by the uniformity of CR, we evaluated the relevant calibration ratios from data using the estimators $Q_{A} = W_{1}/(W_{1}+W_{3})$ and $Q_{B} = W_{1}/(W_{1}+W_{2})$ for the couples of triangular bars shown in Fig. \ref{fig:station}, where $W_{1,2,3} = c_{1,2,3} \times w_{1,2,3}$ are the signal amplitudes of bars 1,2 and 3 scaled by the relative equalization coefficients $c_{i}$. The comparison between the estimators $Q_{A}$ and $Q_{B}$ calculated by using the output signals of bars 1, 2 and 3 and the MC predictions is shown in Fig. \ref{fig:subfig2}. The equalization coefficients have been computed in this way for all the couples of adjacent or faced bars and have been found to differ by few \% from the ones determined using laser pulses (only in one case we obtain a value a little bit less than 15\%)

\begin{figure}[t!]
 \centering
  \includegraphics[width=0.52\textwidth]{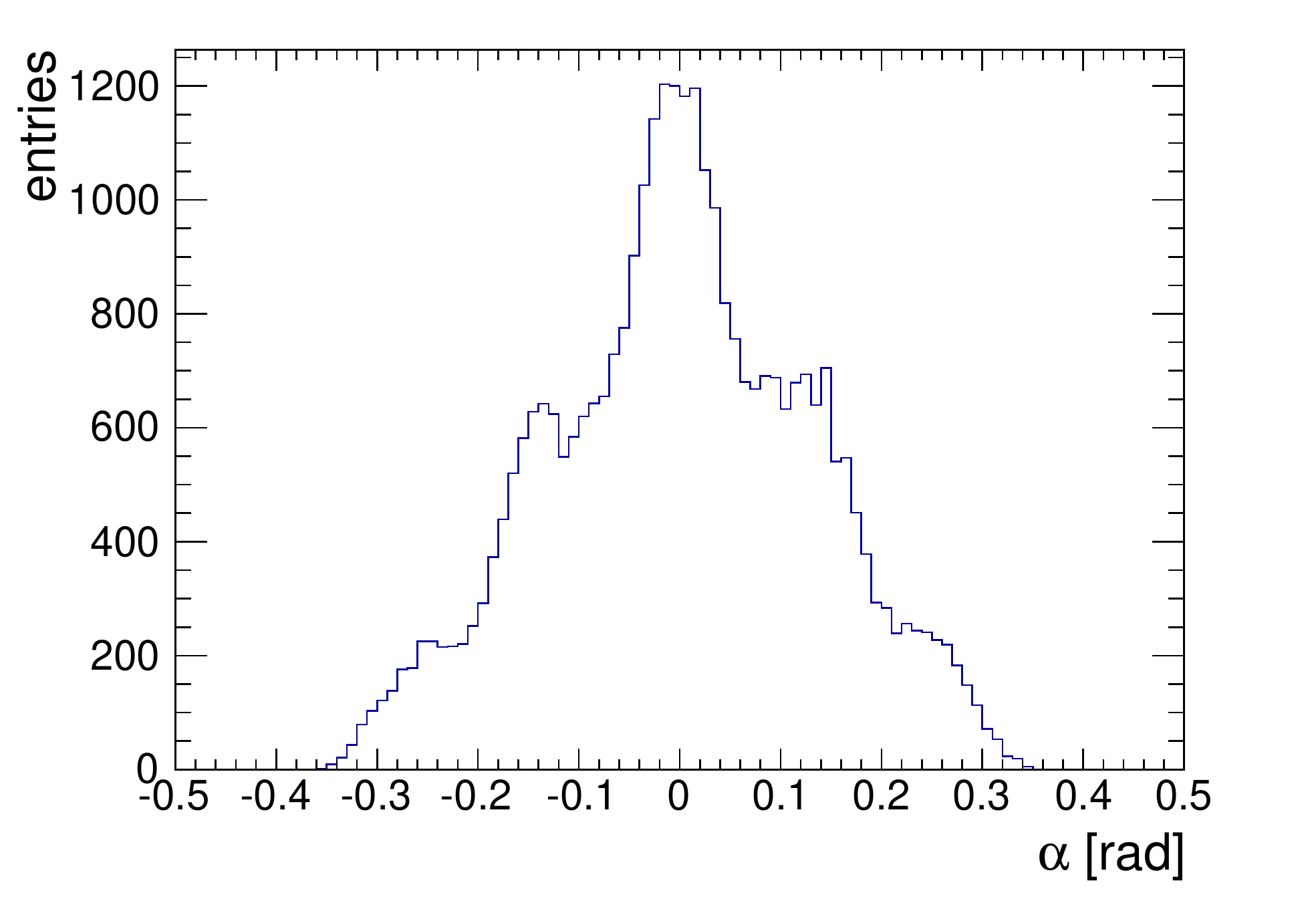}
 \caption{\footnotesize $\alpha$ angle distribution for events with 2 adjacent triangular bars fired in all planes and $\chi^{2} <$ 10.}
\label{fig:slope_adj}
\end{figure}

\begin{figure}[t!]
 \centering
  \includegraphics[width=0.4\textwidth]{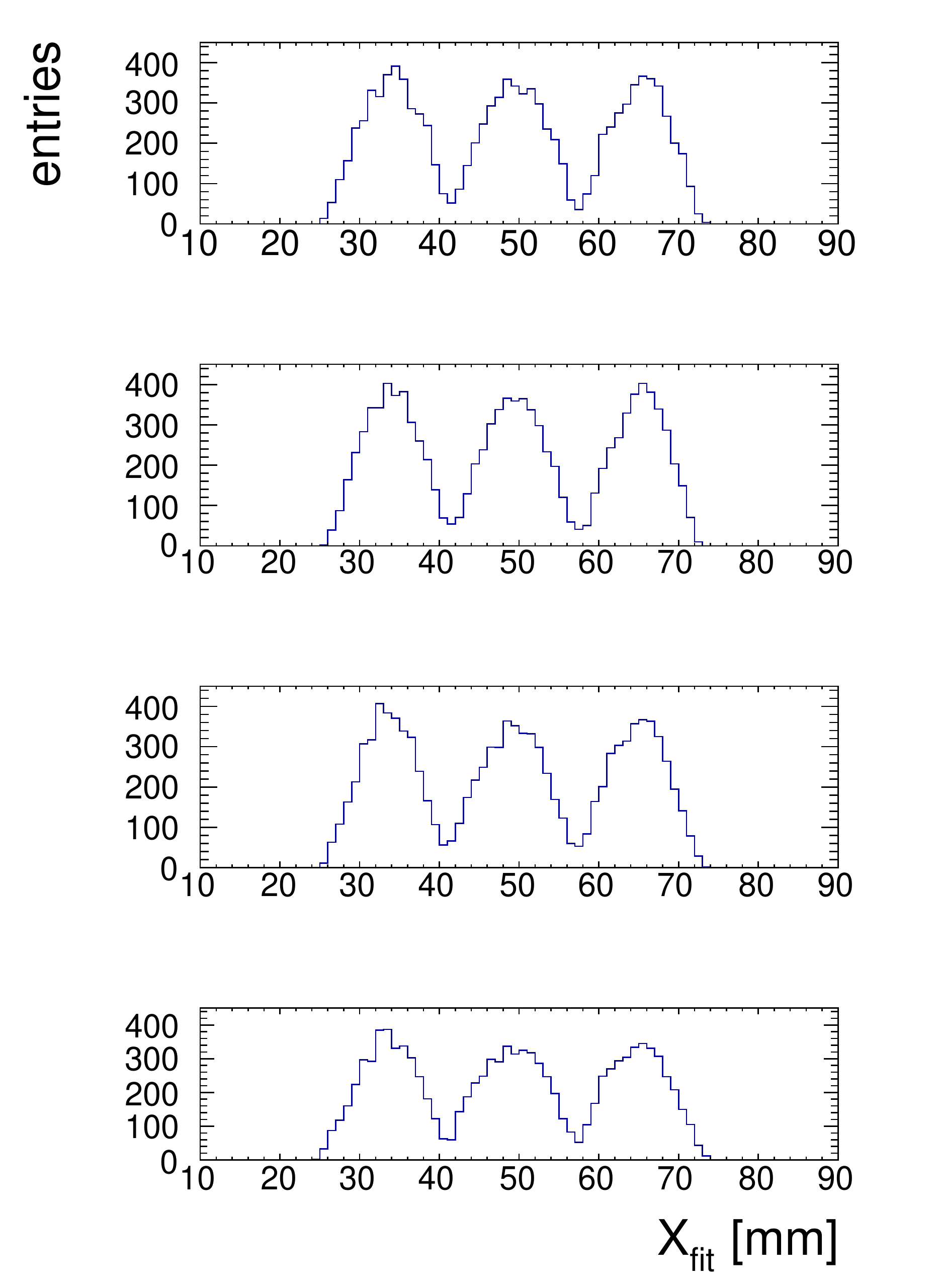}
 \caption{\footnotesize Fitted positions of CR events in each plane of triangular bars for $\chi^{2} <$ 10 and $\alpha <$ 50 mrad. Only tracks that crossed two adjacent triangular bars in each plane were selected. }
\label{fig:pos_fittrackadj_data_cut50mrad}
\end{figure}

\begin{figure}[t!]
 \centering
  \includegraphics[width=0.4\textwidth]{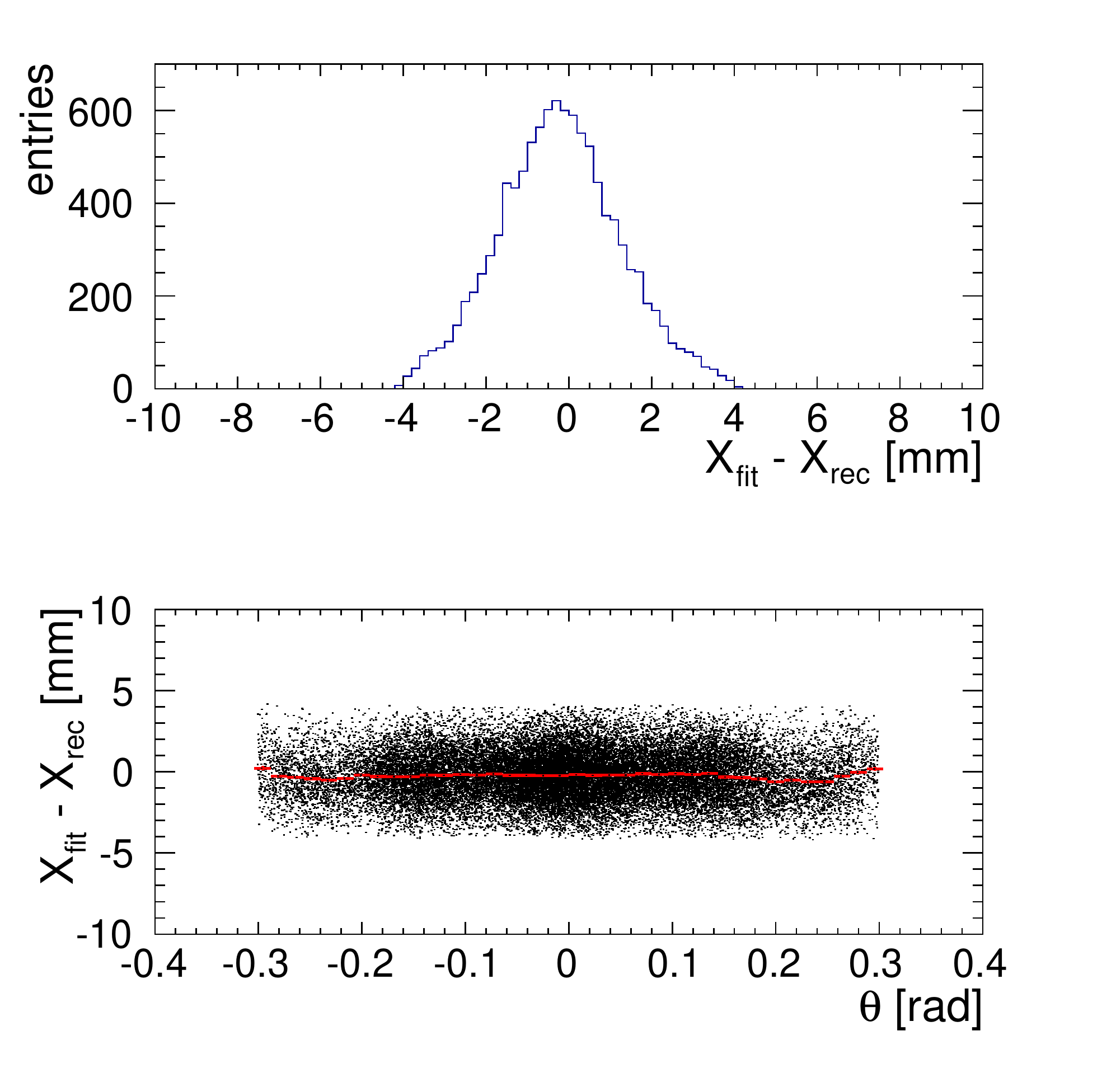}
 \caption{\footnotesize (top) Differences between the reconstructed positions and the projected positions of fitted tracks in the top plane of module 1 for events with 2 adjacent triangular bars fired in each plane for $\chi^{2} <$ 10 and $\alpha <$ 50 mrad. The corresponding spatial resolution is 1.5 mm. (bottom) The residuals between the reconstructed positions and the projected positions of fitted tracks as function of the track inclination $\alpha$.}
\label{fig:residui}
\end{figure}

\subsubsection{Spatial resolution}
 The reconstructed $X$ position of CR events was evaluated for each plane; then a linear fit was performed using an uncertainty of 2 mm on x direction and 1 mm on z direction. The distribution of $\alpha$, the angle from the vertical obtained by the projection of the track in the $x-z$ plane, for events with 2 adjacent bars fired in each plane is reported in Fig. \ref{fig:slope_adj}. For this subsample of events the fitted positions for $\chi^{2} <$  10 and $\alpha <$  50 mrad are shown in Fig. \ref{fig:pos_fittrackadj_data_cut50mrad}. As one can see the number of events collected has the same distribution over each plane, thus implying a uniformity in the efficiency of adjacent bars responses.
 
In order to give an estimate of plane efficiency, in a run one plane was excluded from the trigger and fitted tracks were computed from reconstructed position over 3 planes. Events were selected with only two adjacent bars fired and a cut of $\alpha <$ 50 mrad was applied. Then in the plane excluded from the trigger we selected only events with signals in the corresponding couples of adjacent bars. The efficiency was evaluated as the ratio of the number of selected events to the total number of triggered events. A value of 85\% for this efficiency was found. In order to estimate the efficiency of single triangular bar we computed the ratio of the number of events with a signal in a single bar in the corresponding couple of adjacent bar of the plane excluded to the total number of triggered events. Thus our single bar efficiency resulted to be about 92 \%.

To evaluate the spatial resolution of our test apparatus the differences between reconstructed and fit positions were calculated for each plane. The distribution of these residuals for the top plane of module 1 is shown in Fig. \ref{fig:residui}. In the top panel of the figure only tracks with $\chi^{2} <$ 10 and $\alpha <$ 50 mrad are selected and a spatial resolution of 1.5 mm is obtained. Results for other planes are consistent with each other and ranges from 1.4 to 1.8 mm. 

\begin{figure}[h!]
\centering
\subfloat[][\emph{}]
{\includegraphics[width=.35\textwidth]{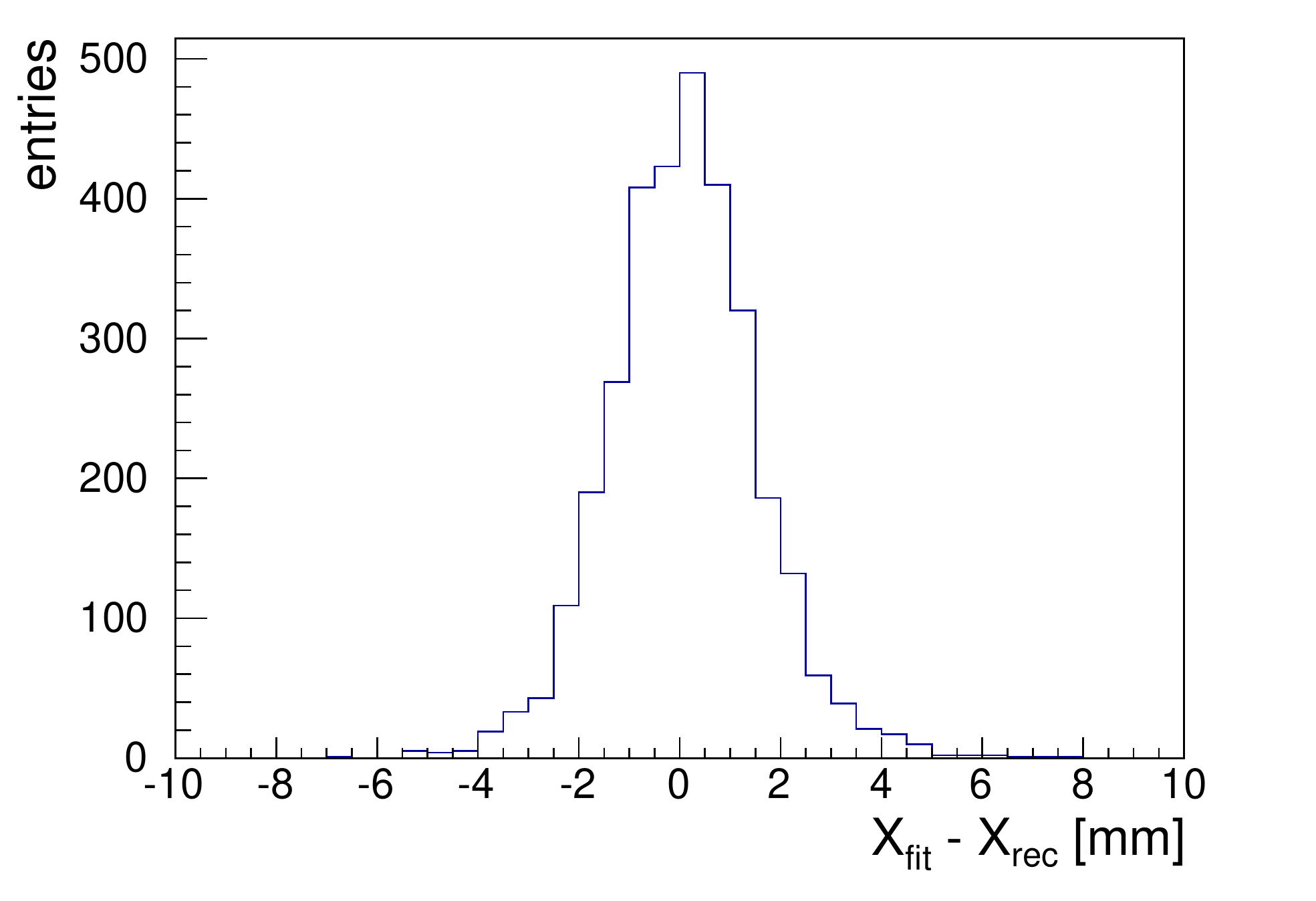}} \quad
\subfloat[][\emph{}]
{\includegraphics[width=.35\textwidth]{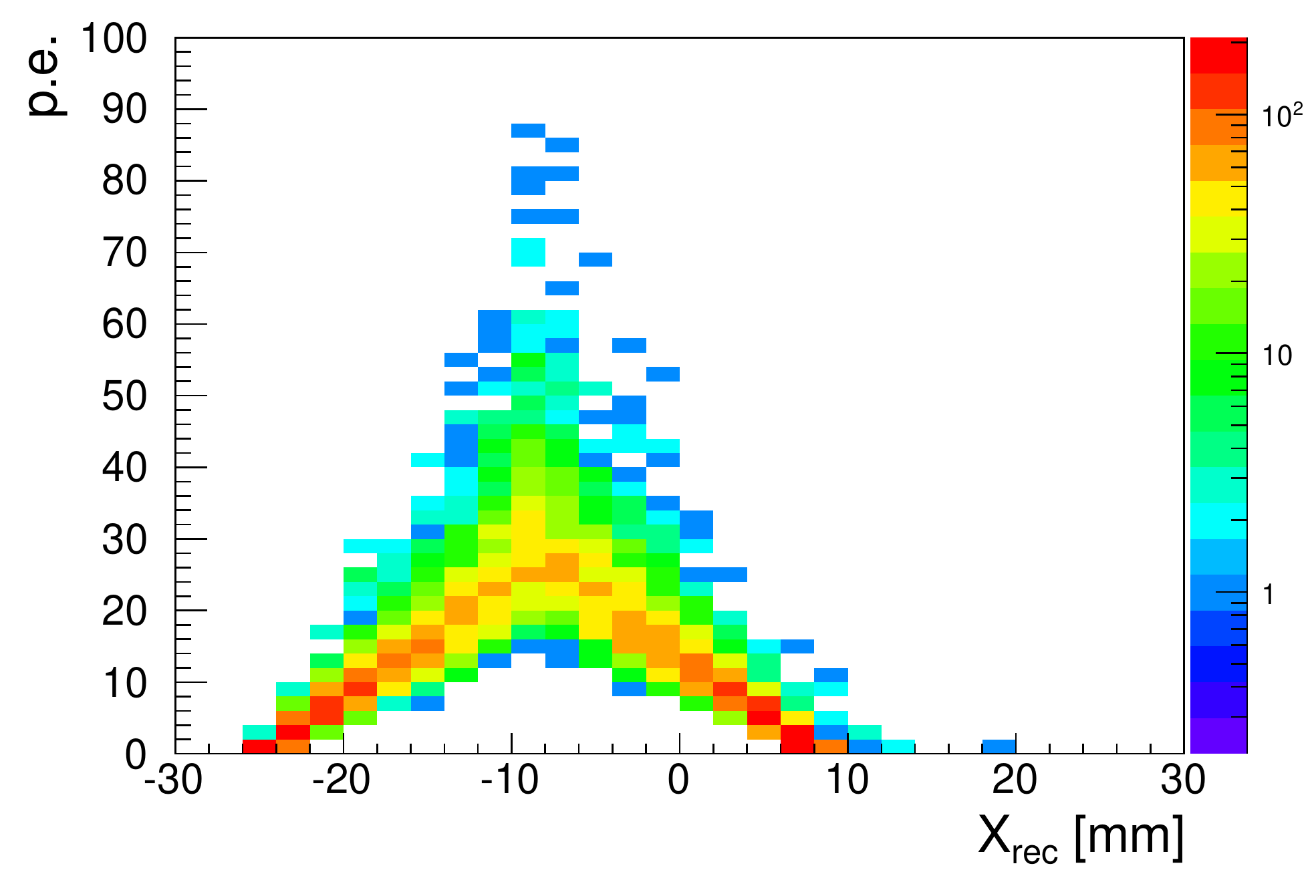}}
\caption{\footnotesize (a) MC distribution of residuals in a plane of triangular bars; (b) MC distribution of the number of collected p.e. in a couple of adjacent bars vs. reconstructed position.}
\label{fig:subfig_MC}
\end{figure}

This result has been compared to the predictions of a more complete MC simulation which takes into account light transport in the bar and the PDE of SiPM. Only muons impinging vertically on the detector plane have been considered. The distribution of the residuals obtained with this MC is shown in Fig. \ref{fig:subfig_MC}a and the number of collected p.e. in a couple of adjacent bars vs reconstructed position is shown in Fig. \ref{fig:subfig_MC}b. A fit with a Gaussian distribution to the former gives a standard deviation of (1.4 $\pm$ 0.1) mm. Thus these two MC predictions are in good agreement with measurements reported in Fig. \ref{fig:residui} and Fig. \ref{fig:signal_posRecp}.


\afterpage{\FloatBarrier}

\section{Conclusions and perspectives}
We developed a detector system made of triangular scintillators bars equipped with SiPMs read in analog mode in order to track low energy charged particles.

Tests with a simple setup have shown that a spatial resolution of 1.8 mm or better is achievable in reconstructing particle tracks. 

This experimental approach will be implemented in a multiplane prototype to be tested with charged beams to set the ultimate spatial accuracy. We expect an improvement in the spatial resolution with a tracking device which foresees triangular bar planes orthogonal to each other, the introduction of a fine control of working condition and a better signal calibration and treatment, as well the use of SiPMs of last generation. 

\section{Acknowledgements}
We gratefully thank A. Candela, D. Sablone and V. Conicella of INFN-LNGS for supplying prototype rectangular bars and front-end boards for their readout. We also want to thank the colleagues F. Zuffa and A. Zucchini of the mechanic's workshop and Pellegrini of the electronic's group of INFN of Bologna for their contribution to this work. We acknowledge A. Montanari and P. Bernardini for very useful discussions and suggestions. Finally we thank L. Stanco and the colleagues of NESSiE project for the encouraging support to this work.





\bibliographystyle{elsarticle-num}

\end{document}